\documentclass[10pt,letterpaper]{article}
\usepackage[T1]{fontenc}
\usepackage{lmodern}
\usepackage[letterpaper,left=0.68in,right=0.68in,top=0.68in,bottom=0.76in,footskip=25pt]{geometry}
\usepackage{graphicx,amsmath,amssymb,amsthm,mathtools}
\usepackage{xcolor,titlesec,caption,booktabs,array}
\usepackage[super,sort&compress]{natbib}

\definecolor{paperlink}{HTML}{CD005C}
\definecolor{papertitle}{HTML}{1E1E23}
\definecolor{papermeta}{HTML}{505058}
\definecolor{paperrule}{HTML}{D2D2D8}
\definecolor{abstractgrey}{HTML}{F7F7F8}

\usepackage[colorlinks=true,linkcolor=paperlink,citecolor=paperlink,urlcolor=paperlink,filecolor=paperlink,pdfborder={0 0 0}]{hyperref}

\titleformat{\section}{\sffamily\bfseries\fontsize{10}{12}\selectfont}{\thesection.}{0.6em}{}
\titleformat{\subsection}{\sffamily\bfseries\fontsize{9}{11}\selectfont}{\thesubsection.}{0.6em}{}
\titlespacing*{\section}{0pt}{9pt}{5pt}
\titlespacing*{\subsection}{0pt}{7pt}{4pt}
\theoremstyle{plain}
\newtheorem{theorem}{Theorem}[section]
\newtheorem{proposition}[theorem]{Proposition}
\newtheorem{lemma}[theorem]{Lemma}
\newtheorem{corollary}[theorem]{Corollary}
\theoremstyle{definition}
\newtheorem{definition}[theorem]{Definition}
\theoremstyle{remark}
\newtheorem{remark}[theorem]{Remark}

\DeclareMathOperator{\Exp}{Exp}
\DeclareMathOperator{\Log}{Log}
\DeclareMathOperator{\Hess}{Hess}
\DeclareMathOperator{\Var}{Var}
\DeclareMathOperator{\Cov}{Cov}
\DeclareMathOperator{\tr}{tr}
\DeclareMathOperator{\diag}{diag}
\DeclareMathOperator{\acosh}{acosh}
\newcommand{\dd}{\mathop{}\!\mathrm{d}}
\newcommand{\C}{\mathbb{C}}
\newcommand{\R}{\mathbb{R}}
\newcommand{\D}{\mathbb{D}}
\newcommand{\HPD}{\mathbb{H}}


\hypersetup{pdftitle={Geodesics and intrinsic statistics of isolated connections in Hermitian positive-definite matrices},
  pdfauthor={Ronald Garc\'ia Reyes and Pedro Antonio Vald\'es-Sosa}}

\providecommand{\doi}[1]{\href{https://doi.org/#1}{doi: \nolinkurl{#1}}}
\newcommand{\paperaffiliation}[2]{%
  \par\noindent\hangindent=1.3em\hangafter=1%
  \makebox[1.3em][l]{\textsuperscript{\textcolor{paperlink}{#1}}}#2\par}
\newcounter{paperalgorithm}
\begin{document}
\pagestyle{plain}
\twocolumn[{\begin{minipage}{\textwidth}

\noindent{\color{papertitle}\sffamily\bfseries\fontsize{20}{23.5}\selectfont
Geodesics and intrinsic statistics of isolated\\[1pt]
connections in Hermitian positive-definite matrices\par}
\vspace{7pt}
\noindent{\sffamily\fontsize{12}{14}\selectfont
Ronald Garc\'ia Reyes\textsuperscript{\textcolor{paperlink}{1,2,3}}\quad\&\quad
Pedro Antonio Vald\'es-Sosa\textsuperscript{\textcolor{paperlink}{1,2,3}}\par}
\vspace{5pt}
{\color{papermeta}\sffamily\fontsize{8}{10}\selectfont
\paperaffiliation{1}{Clinical Hospital of Chengdu Brain Science Institute, University of Electronic
Science and Technology of China, Chengdu 610054, China}
\vspace{2pt}
\paperaffiliation{2}{Neuroinformatics, Cuban Neurosciences Center, 11300 Havana, Cuba}
\vspace{2pt}
\paperaffiliation{3}{China--Cuba Belt and Road Joint Laboratory on Neurotechnology and
Brain--Apparatus Communication, University of Electronic Science and Technology
of China, Chengdu, China}
\vspace{3pt}
\noindent Correspondence: \href{mailto:gaussn0ise29@gmail.com}{gaussn0ise29@gmail.com}\par}
\vspace{8pt}
\setlength{\fboxsep}{7.5pt}
\noindent\colorbox{abstractgrey}{\begin{minipage}{\dimexpr\textwidth-15pt\relax}
{\sffamily\bfseries\fontsize{8}{10}\selectfont Abstract\par}
\vspace{3pt}
{\fontsize{8}{9.8}\selectfont
Quantifying change in a single covariance, coherence or spectral connection
requires a distance that respects the positive-definiteness constraint and holds
the surrounding entries fixed. We characterize the Riemannian geometry induced
by the affine-invariant metric when one conjugate off-diagonal pair of a
Hermitian positive-definite matrix varies under this constraint. A fixed Schur
congruence maps the admissible set to an open complex disk and reduces its
metric to a two-dimensional form independent of the ambient matrix dimension.
The disk is complete, with Gaussian curvature in $(-1/4,0]$, and has unique
minimizing geodesics and globally defined logarithms. Its intrinsic distance
strictly exceeds the ambient affine-invariant distance between distinct points.
Conservation of angular momentum reduces the geodesic boundary-value problem
to a single scalar equation with elementary angular and length integrals,
enabling distance and logarithm evaluation without shooting. For a declared
reference distribution with a finite second intrinsic moment, we establish a
unique Fr\'echet centre and consistency of its empirical estimate, together with
a sandwich central limit theorem under the stated regularity conditions.
Tangent covariance standardization yields moment identities, and exchangeable
rank calibration supplies finite-sample score guarantees. These results provide
a geometric and statistical framework for isolated connections on a fixed
complement, with explicit limits on extension to varying complements.\par}
\vspace{6pt}
{\sffamily\fontsize{8}{10}\selectfont\textbf{Keywords:} intrinsic geometry;
Hermitian positive-definite matrices; Schur complement; Fr\'echet mean;
nonpositive curvature; conformal calibration\par}
\vspace{2pt}
{\sffamily\fontsize{8}{10}\selectfont\textbf{2020 Mathematics Subject
Classification:} 53C21 (primary); 62H99, 53B12 (secondary)\par}
\par\vspace{6pt}
\noindent\includegraphics[width=\linewidth]{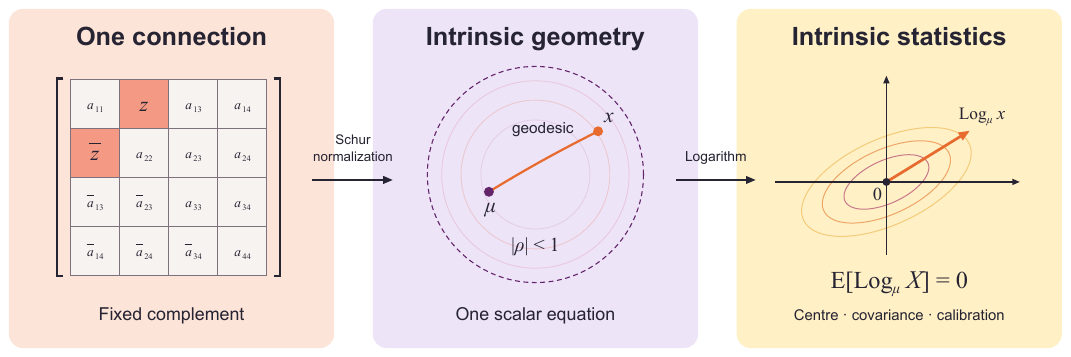}\par
\end{minipage}}
\vspace{3pt}\noindent{\color{paperrule}\rule{\textwidth}{0.4pt}}\par
\vspace{9pt}

\end{minipage}}]

\section{Introduction}
\label{sec:intro}

An entry of a covariance, coherence or spectral-density matrix does not vary
freely. Its admissible values are bounded by the entries around it, and a
displacement that looks modest in isolation can make the whole matrix singular.
For a labelled connection the natural question is therefore not how far it can
move in some ambient geometry, but how far it can move while every other entry is
held fixed. That question fixes both a domain and a family of admissible paths,
and the answer depends on the constraint as well as on the chosen matrix metric.

A fixed-complement model describes, for example, a controlled sensitivity analysis in which one connection is perturbed while the measured diagonal and all other connections are held at their reference values. The resulting distance quantifies admissible change along that slice. Statistical inference across observations with varying surrounding entries requires an additional model, as discussed in Section~\ref{sec:calibration}.

Other analyses of connectivity treat the matrix as a whole. For example, intrinsic local polynomial regression models how a positive-definite matrix varies with covariates~\cite{GarciaReyesILPR}, while the Janashia--Lagvilava algorithm studied by Wang et al.\ factors a matrix-valued spectrum~\cite{WangJL}. Here the surrounding entries instead serve as constraints on a single connection.

Geometry for correlation matrices already offers answers to adjacent questions.
David and Gu formulated correlation geometry as a quotient by positive diagonal
congruence, including the complex Hermitian case~\cite{G01}. In the real
two-variable case this yields a distance proportional to a difference of Fisher
transforms~\cite{G03}, and transformed means and dispersions have been used for
anomaly analysis~\cite{G02}. Later work established alternative correlation
geometries and identified curvature restrictions on the full quotient-affine
correlation manifold~\cite{G04,G05}. A quotient removes scale directions;
requiring the diagonal and the complementary block to stay fixed instead
restricts admissible motion. These are different constraints and they produce
different metrics.

For covariance matrices the Schur-normalized entry is the classical partial
correlation of the two variables after linear adjustment for the complement.
After a fixed congruence our surface is exactly the unit-diagonal
two-by-two Hermitian correlation family equipped with the restriction of the
ambient metric. Results for that induced submanifold therefore apply verbatim.
The quotient construction instead minimizes over diagonal rescaling velocities.
On the real diameter the squared radial coefficients are respectively
$2(1+r^2)/(1-r^2)^2$ and $2/(1-r^2)^2$. The distinction also matters for
comparisons with totally geodesic SPD submanifolds~\cite{G07}, since the
ambient geodesic leaves our slice (Section~\ref{sec:comparison}).

Metric choice remains consequential. Log-Euclidean and log-Cholesky
constructions supply different coordinate geometries~\cite{G16,G20,G21},
and statistical models on positive-definite manifolds depend on that
choice~\cite{G26,G27,G28}. Here the metric and path constraint are fixed
before the distance is defined.

Fixed-entry positive-definite completion and partial-correlation
parameterizations supply the algebraic background~\cite{G32,G33,G34,G35}. We use
the Schur complement to isolate one conjugate pair and express its feasible values
as a normalized disk. Schur feasibility, correlation geometry, Fr\'echet
averaging~\cite{S03,S09} and covariance whitening~\cite{S14} are established
ingredients; the contribution developed here is the explicit fixed-complement reduction,
the elementary geodesic-leg solution with its branch conditions, and the
population and computational consequences on that induced surface.

Three features of the induced surface are worth stating at the outset. First, one
congruence that is constant across the leaf removes the complement from the
tangent energy, so the geometry is two-dimensional in the complex case and
one-dimensional in the real case, independently of the ambient matrix dimension.
Second, the complex surface is complete and nonpositively curved, with curvature
varying between zero and $-1/4$; the Cartan--Hadamard theorem then gives unique
minimizing geodesics and global logarithms, and the resulting convexity of the
squared distance, together with a finite second moment, gives a unique empirical
Fr\'echet centre. Third, the real
restriction has an elementary signed arc coordinate, and that sign is exactly what
a squared joint score cannot retain.

The geodesic problem on the disk reduces to a single scalar equation. Angular
momentum is conserved, so every pair of points determines a Clairaut constant
through either a monotone branch or a branch with one radial turning point. We
show that both legs of that reduction are exactly integrable, which turns the
numerical problem into one bounded root find followed by an evaluation, and we
give the branch conditions and the limiting cases at the origin, on a diameter
and at equal radii. The elementary legs avoid quadrature at a moving turning point, where
a fixed-node rule can misjudge an objective difference. Stable limits and
precision refinement are therefore part of the evaluation procedure.

Statistical standardization requires a probability law in addition to a metric.
Intrinsic means and covariance-weighted logarithms are established
tools~\cite{S03}; whitening fixes the covariance while leaving rotational freedom
in the interpretation of components~\cite{S14}. We apply these tools to a
declared fixed-complement law, separate individual dispersion from uncertainty in
the estimated centre~\cite{S02}, and give a conservative rank for a score frozen
after fitting under exchangeability~\cite{S10}. The limit theory we need is the
Fréchet-mean central limit theorem and its refinements on manifolds with and
without cut loci~\cite{S04,S05,S06,S07,S20}, together with convergence rates for
empirical barycenters in metric spaces~\cite{S08}; regression and analysis of
variance for manifold-valued responses provide the natural population
frameworks~\cite{S15,S16}. Common normalized geometry neither identifies a common
distribution nor supplies conditional calibration for arbitrary pooled
complements~\cite{S12}, and the conformal literature makes the finite-sample
guarantee and its exchangeability requirement precise~\cite{S11,S13,S17,S18,S19}.

Section~\ref{sec:surface} constructs the surface and proves completeness and the
curvature bound. Section~\ref{sec:reference} develops the reference population,
the covariance standardization and the rank cases.
Section~\ref{sec:comparison} compares the constrained distance with ambient and
quotient constructions, and Section~\ref{sec:invariance} records the invariances
of the declared slice. Section~\ref{sec:geodesics} solves the geodesic problem
and defines logarithms and parallel transport.
Section~\ref{sec:calibration} treats estimation, calibration and
independent-replicate inference. Section~\ref{sec:numerical} reports the
numerical validation and Section~\ref{sec:discussion} discusses scope and limits.
The appendices collect the ambient support reduction, the varying-complement
derivative, the stable evaluation of the arc coordinate and the moment
and derivative conditions used by the statistical limit theory.

  \begin{figure*}[t]%
  \centering
  \includegraphics[width=\textwidth]{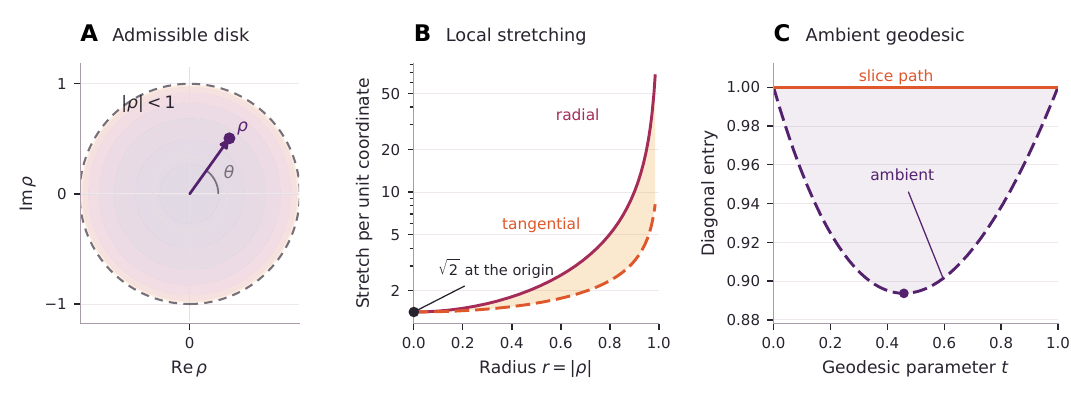}%
  \caption{\textbf{An isolated connection has its own path geometry.} (A) The admissible disk of normalized connections; the heat contours show intrinsic distance from the origin. Only the selected conjugate pair varies. (B) Radial and tangential stretch agree at the origin and diverge toward the boundary. (C) The ambient geodesic changes the diagonal entries, whereas a slice path keeps them fixed.}\label{fig:figure_1.pdf}%
  \end{figure*}

\section{An isolated connection as a Riemannian surface}
\label{sec:surface}

\subsection{The declared slice and its reduction}

A connection is defined in a declared matrix frame. Throughout, $n\ge 2$,
$A$ is Hermitian positive definite (HPD), and only the entries $A_{12}=z$ and
$A_{21}=\bar z$ are permitted to vary. After a simultaneous row and column
permutation that places the selected pair first, write
\begin{equation}
 A(z)=\begin{pmatrix}B(z)&E\\ E^*&C\end{pmatrix},
 \qquad
 B(z)=\begin{pmatrix}a&z\\ \bar z&b\end{pmatrix},
 \label{eq:block}
\end{equation}
with $C\in\C^{(n-2)\times(n-2)}$ and $E\in\C^{2\times(n-2)}$.
For $n=2$ the blocks $E$ and $C$ are absent and no empty inverse is used.

The metric convention throughout is the affine-invariant Riemannian metric
\begin{equation}
 g_A(U,V)=\operatorname{Re}\tr\!\left(A^{-1}UA^{-1}V\right),
 \label{eq:airm}
\end{equation}
on Hermitian tangent vectors with real scalar coefficients. A complex connection
therefore carries two real degrees of freedom. The adjoint $*$ denotes conjugate
transpose, and the norm associated with Eq.~\eqref{eq:airm} is its positive
square root. Entries may carry covariance units; $\rho$, lengths under
Eq.~\eqref{eq:airm} and standardized scores are dimensionless.

\begin{definition}[Fixed-complement slice]
\label{def:slice}
For $n>2$, fix $a,b\in\R$, $E$ and $C\succ0$, and let $H=EC^{-1}E^*$, with
$\alpha=a-H_{11}$, $\beta=b-H_{22}$ and $\gamma=H_{12}$. Assume
$\alpha,\beta>0$. The slice is the set
$\mathcal S=\{A(z): z\in\C,\ A(z)\succ0\}$ together with the restriction of
Eq.~\eqref{eq:airm}. The normalized connection is
$\rho=(z-\gamma)/\sqrt{\alpha\beta}$. For $n=2$, set $H=0$ and $\gamma=0$, so $\alpha=a$ and $\beta=b$.
\end{definition}

With dense inputs one Cholesky factorization of $C$ costs $O((n-2)^3)$
operations. Two triangular solves and their Gram matrix give $H$ in a
further $O((n-2)^2)$ operations. The same $H$ serves all endpoints on the
slice. Each subsequent normalization has constant cost in $n$, while the
scalar geodesic solver also depends on endpoint conditioning and precision.
Small Schur pivots amplify relative errors in the normalization. Singular
positive-semidefinite estimates lie outside this model. Regularization changes
the matrix and thereby the connection being scored.

\begin{theorem}[Feasibility and dimension-independent reduction]
\label{thm:reduction}
Under Definition~\ref{def:slice}, $A(z)\succ0$ if and only if $|\rho|<1$.
Moreover a single congruence, independent of $\rho$, maps $A(z)$ to
$\diag(R(\rho),I_{n-2})$ with
\begin{equation}
 R(\rho)=\begin{pmatrix}1&\rho\\ \bar\rho&1\end{pmatrix},
 \label{eq:R}
\end{equation}
and carries the metric~\eqref{eq:airm} on the slice to that of $R(\rho)$ alone.
The complementary block contributes exactly zero to the induced metric.
\end{theorem}

\begin{proof}
For $n>2$ multiply on the left by
$L=\begin{psmallmatrix}I_2&-EC^{-1}\\0&I_{n-2}\end{psmallmatrix}$ and on the right
by $L^*$. Block multiplication gives
$L\,A(z)\,L^*=\diag(S(z),C)$ with
\begin{equation}
 S(z)=B(z)-EC^{-1}E^*=
 \begin{pmatrix}\alpha&z-\gamma\\ \bar z-\bar\gamma&\beta\end{pmatrix}.
 \label{eq:schur}
\end{equation}
Since $L$ is invertible, Sylvester's law of inertia gives
$A(z)\succ0\iff C\succ0$ and $S(z)\succ0$. For the $2\times2$ matrix
$S(z)$, positive definiteness is equivalent to $\alpha>0$ and
$\det S=\alpha\beta-|z-\gamma|^2>0$; together these imply $\beta>0$.
Writing $z-\gamma=\rho\sqrt{\alpha\beta}$ turns the determinant condition into
$|\rho|<1$, proving the first claim.

For the second, let $D=\diag(\alpha,\beta)$ and set
$T=\diag(D^{-1/2},C^{-1/2})L$. Then
$T\,A(z)\,T^*=\diag(R(\rho),I_{n-2})$, because
$D^{-1/2}S(z)D^{-1/2}=R(\rho)$ and $C^{-1/2}CC^{-1/2}=I_{n-2}$.

It remains to show that $T$ is a tangent isometry for~\eqref{eq:airm}. For any
invertible $T$, $(TAT^*)^{-1}=T^{-*}A^{-1}T^{-1}$. Substituting
$U\mapsto TUT^*$ and $V\mapsto TVT^*$ gives
\[
 \tr\bigl(T^{-*}A^{-1}T^{-1}TUT^*\,T^{-*}A^{-1}T^{-1}TVT^*\bigr)
\]
\[
 \phantom{\tr\bigl(}=\tr(A^{-1}UA^{-1}V)
\]
by cyclicity of the trace, so Eq.~\eqref{eq:airm} is invariant under congruence.
Applying this to $T$ and restricting to variations $\dd A$ supported on the
selected pair leaves exactly
$\operatorname{Re}\tr(R^{-1}\dd R\,R^{-1}\dd R)$. For $n=2$ the same argument
applies with $T=D^{-1/2}$ only.
\end{proof}

The set $\mathcal S$ is a submanifold of the HPD manifold, and the question of
which submanifolds of the positive-definite cone are totally geodesic has been
studied in its own right~\cite{G07}; the slice here is not one of them in
general, which is exactly why its induced geodesics differ from ambient ones.
If no positive complement or no positive $\alpha,\beta$ exists, the slice is
empty; a disk written formally over a non-positive block is not the geometry of
the original matrices. On a non-empty slice the normalization is invertible, so
distinct normalized connections are identifiable once the frame and the
complementary entries have been declared. Figure~\ref{fig:figure_1.pdf} draws
the admissible disk schematically, and evaluates the induced stretch and an
ambient path that leaves the slice.

\subsection{The induced metric}

\begin{theorem}[The induced complex metric]
\label{thm:metric}
Identify $\rho=x+\mathrm iy$ with $q=(x,y)^\top$ in the open unit disk $\D$ and
put $r=|\rho|$. The metric induced by Eq.~\eqref{eq:airm} is
\begin{equation}
 G(q)=\frac{2}{1-r^2}I_2+\frac{4}{(1-r^2)^2}\,qq^\top .
 \label{eq:cartmetric}
\end{equation}
In polar coordinates $\rho=r\mathrm e^{\mathrm i\theta}$, away from the origin,
\begin{equation}
 \dd s^2=a(r)\,\dd r^2+b(r)\,\dd\theta^2,
 \label{eq:polar}
\end{equation}
where
\begin{equation}
 a(r)=\frac{2(1+r^2)}{(1-r^2)^2},\qquad
 b(r)=\frac{2r^2}{1-r^2}.
 \label{eq:polarcoef}
\end{equation}
\end{theorem}

\begin{proof}
Write $\Delta=1-|\rho|^2$ and $w=\dd\rho$. Inverting $R(\rho)$ gives
$R^{-1}=\Delta^{-1}\begin{psmallmatrix}1&-\rho\\-\bar\rho&1\end{psmallmatrix}$
and $\dd R=\begin{psmallmatrix}0&w\\\bar w&0\end{psmallmatrix}$, hence
\[
 R^{-1}\dd R=\Delta^{-1}
 \begin{pmatrix}-\rho\bar w&w\\ \bar w&-\bar\rho w\end{pmatrix}.
\]
Taking the trace of the square,
\[
 g(w,w)=\Delta^{-2}\left(\rho^2\bar w^2+\bar\rho^2w^2+2|w|^2\right)
\]
\[
 \phantom{g(w,w)}=\Delta^{-2}\left(2|w|^2+2\operatorname{Re}(\bar\rho^2w^2)\right).
\]
Since $\bar\rho^2w^2=(\bar\rho w)^2$ and
$\operatorname{Re}[(\bar\rho w)^2]=2[\operatorname{Re}(\bar\rho w)]^2-r^2|w|^2$,
\[
 g(w,w)=\frac{2|w|^2}{\Delta}+\frac{4[\operatorname{Re}(\bar\rho w)]^2}{\Delta^2}.
\]
Now $|w|^2=\dd x^2+\dd y^2$ and
$\operatorname{Re}(\bar\rho w)=x\,\dd x+y\,\dd y=q\cdot\dd q$, which gives
Eq.~\eqref{eq:cartmetric}. Polarizing recovers the bilinear form, including its
mixed terms. In polar coordinates $|w|^2=\dd r^2+r^2\dd\theta^2$ and
$\operatorname{Re}(\bar\rho w)=r\,\dd r$, so
$g=\bigl(2\Delta^{-1}+4r^2\Delta^{-2}\bigr)\dd r^2+2r^2\Delta^{-1}\dd\theta^2$,
and $2\Delta+4r^2=2(1+r^2)$ yields Eq.~\eqref{eq:polar}.
\end{proof}

At the origin $G(0)=2I_2$ and the Cartesian coefficients are analytic. The
same matrix is the ordinary Cartesian Hessian of the logarithmic barrier
$-\log(1-|q|^2)$, as direct differentiation verifies. This supplies a
second algebraic characterization of the induced metric.
At the origin the
vanishing polar angular coefficient $b(r)\to0$ at $r=0$ is a coordinate artefact,
not a degeneracy of the metric. In particular, extending a signed arc coordinate
independently to the two Cartesian directions would produce a different metric;
the correct extension is the one constructed next.

\subsection{The real arc coordinate}

On the real slice $\rho\in(-1,1)$ define
\begin{equation}
 \begin{aligned}
 \psi(\rho)&=\sqrt2\int_0^\rho\frac{\sqrt{1+t^2}}{1-t^2}\,\dd t\\
 &=2\operatorname{atanh}\frac{\sqrt2\,\rho}{\sqrt{1+\rho^2}}
  -\sqrt2\operatorname{asinh}\rho .
 \end{aligned}
 \label{eq:psi}
\end{equation}

\begin{proposition}[Properties of the arc coordinate]
\label{prop:psi}
The function $\psi$ is real analytic and odd on $(-1,1)$, satisfies
$\psi'(r)=\sqrt2\sqrt{1+r^2}/(1-r^2)$, is strictly increasing, and maps
$(-1,1)$ onto $\R$. Its Taylor expansion at the origin is
\begin{equation}
 \psi(\rho)=\sqrt2\left(\rho+\tfrac12\rho^3+\tfrac{11}{40}\rho^5+O(\rho^7)\right).
 \label{eq:psiseries}
\end{equation}
The signed arc coordinate is $\psi(\rho)$, and the length from $0$ to $\rho$ is $|\psi(\rho)|$. Thus
$\psi$ is an isometry onto $\R$ with the Euclidean metric.
\end{proposition}

\begin{proof}
Differentiating $t=\sinh u$ gives $\dd t=\cosh u\,\dd u$ and
$\sqrt{1+t^2}=\sqrt{1+\sinh^2u}=\cosh u$, so
$\sqrt{1+t^2}\,\dd t=\cosh^2u\,\dd u=(1+\sinh^2u)\,\dd u$. With
$1-t^2=1-\sinh^2u$ the integrand becomes
$(1+\sinh^2u)/(1-\sinh^2u)$. Setting $v=\tanh u$ in the remaining
$\dd u/(1-\sinh^2u)$ integral gives
$\int \dd u/(1-\sinh^2u)=\operatorname{atanh}(\sqrt2 v)/\sqrt2$. Evaluating both
antiderivatives and using $\sqrt2v=\sqrt2\rho/\sqrt{1+\rho^2}$ produces
Eq.~\eqref{eq:psi}; both antiderivatives vanish at $\rho=0$, and the argument of
$\operatorname{atanh}$ lies strictly between $-1$ and $1$ because
$2\rho^2<1+\rho^2$ for $|\rho|<1$. The derivative follows by differentiating the
integrand, and it is positive on $(-1,1)$, so $\psi$ is strictly increasing.
As $\rho\uparrow1$ the second term stays bounded while
$\operatorname{atanh}\to\infty$, so $\psi\to+\infty$; oddness gives the other
end. Continuity and strict monotonicity then give surjectivity onto $\R$.
Expanding each term about $\rho=0$ yields Eq.~\eqref{eq:psiseries}.
\end{proof}

For the complex disk define $s=\psi(r)\ge0$ and the warped-product profile
$f(s)=\sqrt2\,r/\sqrt{1-r^2}$ with $r=\psi^{-1}(s)$. Then
Eq.~\eqref{eq:polar} reads $\dd s^2+f(s)^2\dd\theta^2$.

\begin{theorem}[Curvature, completeness and unique logarithms]
\label{thm:hadamard}
The induced surface $(\D,g)$ is complete, simply connected and has nonpositive
Gaussian curvature
\begin{equation}
 K(s)=-\frac{f''(s)}{f(s)}=-\frac{r^2}{(1+r^2)^2}\in\left(-\tfrac14,0\right],
 \label{eq:curvature}
\end{equation}
with $K(0)=0$ attained only at the origin and $K\to-1/4$ as $r\to1$.
Consequently, for every $p\in\D$ the exponential map
$\Exp_p:T_p\D\to\D$ is a smooth global diffeomorphism and $\Log_p$ is its
inverse. For points on one radial ray the distance is
$|\psi(r_2)-\psi(r_1)|$; for opposite rays it is $\psi(r_1)+\psi(r_2)$; for real
points it is $|\psi(\rho_2)-\psi(\rho_1)|$.
\end{theorem}

\begin{proof}
\emph{Curvature.} With $f(r)=\sqrt2 r(1-r^2)^{-1/2}$ and
$\psi'(r)=\sqrt2\sqrt{1+r^2}/(1-r^2)$,
\begin{gather*}
 f'(s)=\frac{f'(r)}{\psi'(r)}=\frac{1}{\sqrt{1-r^4}},\\
 f''(s)=\frac{1}{\psi'(r)}\frac{\dd}{\dd r}\frac{1}{\sqrt{1-r^4}}
 =\frac{\sqrt2\,r^3}{\sqrt{1-r^2}\,(1+r^2)^2}.
\end{gather*}
For a warped product $\dd s^2+f(s)^2\dd\theta^2$ the only non-vanishing
Christoffel symbols are $\Gamma^s_{\theta\theta}=-ff'$ and
$\Gamma^\theta_{s\theta}=f'/f$, and $R^s{}_{\theta s\theta}=-ff''$, so
$K=-f''/f=-r^2/(1+r^2)^2$. The Cartesian coefficients of $G$ are analytic at the
origin, so this expression extends continuously with $K(0)=0$. Its magnitude
increases strictly on $[0,1)$ and tends to $1/4$.

\emph{Completeness.} The radial length inequality gives
$d(o,x)=\psi(|x|)$: every curve has length at least the radial-coordinate
change, and the radial segment attains it. Let $\gamma:[0,L)\to\D$ be a
curve of finite length $L$.
Since $g(\dot\gamma,\dot\gamma)\ge\dot s^2$ for the radius coordinate
$s(t)=\psi(|\gamma(t)|)$,
\[
 L\ \ge\ \int_0^L|\dot s|\,\dd t\ \ge\ \bigl|\lim_{t\to L}s(t)-s(0)\bigr|,
\]
so a curve of finite length has bounded $s$. Because $\psi(r)\to+\infty$ as
$r\uparrow1$, bounded $s$ confines the curve to a closed Euclidean disk of radius
$R<1$. For completeness, fix a metric-Cauchy sequence. The triangle inequality
gives $d(o,x_n)\le d(o,x_1)+d(x_1,x_n)$, which is bounded, and
$s_n=\psi(|x_n|)=d(o,x_n)$, so the sequence lies in a Euclidean disk of radius
$R<1$. On that disk
$G\preceq M I$ for some finite $M$. The global bound $G\succeq2I$ gives
$d(x,y)\ge\sqrt2\,|x-y|$, so the sequence is Euclidean-Cauchy and converges
to $x$ with $|x|\le R<1$. The straight segment stays in that disk; its
length is at most $\sqrt M\,|x_n-x|$, proving metric convergence. Hence $(\D,g)$ is complete. The disk is simply
connected, and $K\le0$ was proved above; the Cartan--Hadamard theorem then gives
that $\Exp_p$ is a global diffeomorphism for every $p$, and $\Log_p$ its smooth
inverse. Uniqueness of minimizing geodesics follows as well.

\emph{Distances.} A radial curve has constant $\theta$ and satisfies the geodesic
equations in $(s,\theta)$; its length is the variation of $s$, and no shorter
curve has the same endpoints because any curve has length at least the absolute
variation of $s$. A diameter extends smoothly through the origin in Cartesian
coordinates, so its two halves form a single geodesic; uniqueness makes it
minimizing, and its length is $\psi(r_1)+\psi(r_2)$. The real case is the
restriction $\theta\in\{0,\pi\}$.
\end{proof}

\begin{figure*}[t]%
  \centering
  \includegraphics[width=\textwidth]{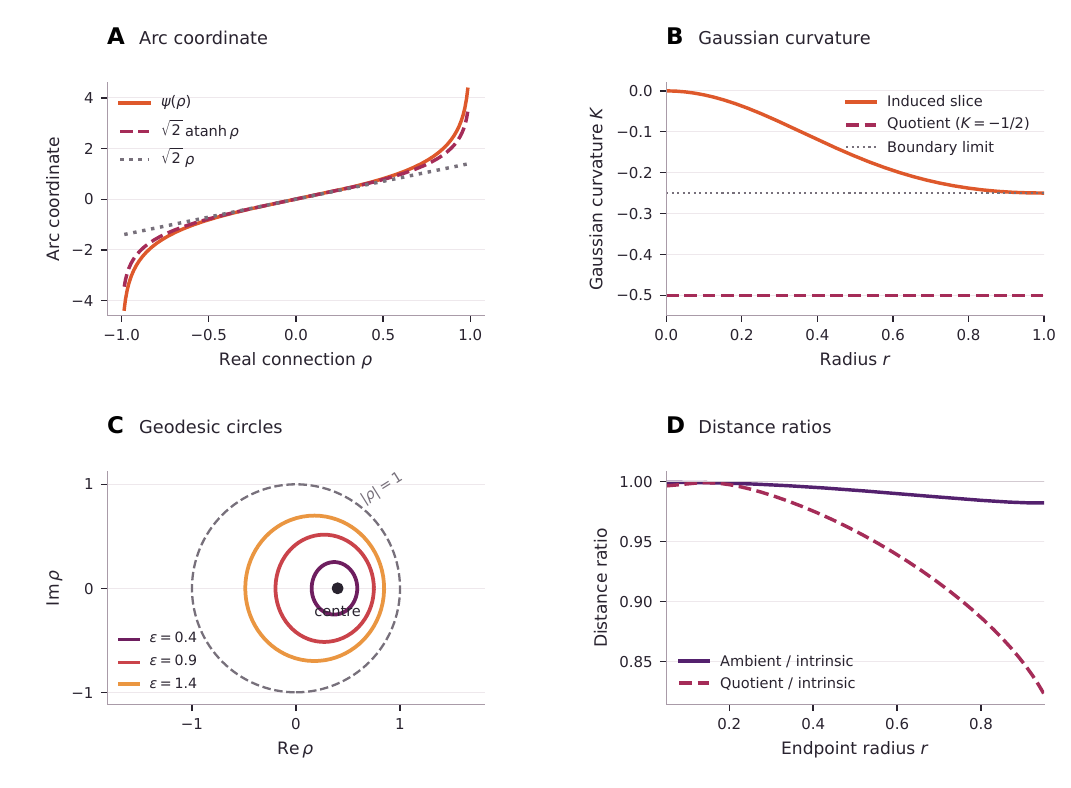}%
  \caption{\textbf{Arc coordinates and variable curvature distinguish the connection disk.} (A) The signed intrinsic arc coordinate, its quotient counterpart and its linearization. (B) Gaussian curvature of the disk, compared with the constant curvature of the quotient and the disk boundary limit. (C) Geodesic circles about an off-centre point, colored by intrinsic radius. (D) Ambient and quotient distances relative to intrinsic distance for a fixed start and varying endpoint.}\label{fig:figure_2.pdf}%
  \end{figure*}

Theorem~\ref{thm:hadamard} derives global existence and uniqueness from
completeness and curvature. General geodesics carry both magnitude and phase; their
computational characterization is developed in Section~\ref{sec:geodesics}.
Figure~\ref{fig:figure_2.pdf} displays the arc coordinate, the curvature, the
numerically integrated geodesic circles and the distance ratios computed from the scalar formulas.

\section{Reference populations and intrinsic standardization}
\label{sec:reference}

A controlled perturbation traces a deterministic path with every complementary
entry held fixed. A random law on that slice additionally specifies how
the selected pair varies. The statistical results below concern this second
setting. Sample covariance and cross-spectral matrices generally fluctuate
in all their blocks. For those estimators Appendix~\ref{app:varying} propagates
a specified joint perturbation through the normalization. Applying a reference
law or a calibration rank across subjects would additionally require a sampling
unit, a model for the complement and an exchangeable null design.

A reference law $P$ is a probability measure on one declared fixed slice,
transported to $\D$ by $\rho$. The same metric on normalized slices does not
imply the same probability law, and no probability statement below is derived
from the geometry alone. Throughout this section $o$ denotes a fixed base point
and $d$ the intrinsic distance of Theorem~\ref{thm:hadamard}. We assume
\begin{equation}
 \mathbb E_P\,d(o,X)^2<\infty,
 \label{eq:secondmoment}
\end{equation}
which holds for one base point if and only if it holds for every other, by the
triangle inequality. Define $F(p)=\tfrac12\mathbb E_P\,d(p,X)^2$.

\begin{theorem}[Centre, dispersion and mean logarithm]
\label{thm:center}
Under Eq.~\eqref{eq:secondmoment} the function $F$ attains its minimum at exactly
one point $\mu\in\D$. That point satisfies the first-order condition
\begin{equation}
 \mathbb E_P\Log_\mu X=0,
 \label{eq:firstorder}
\end{equation}
its dispersion is $D=\mathbb E_P\,d(\mu,X)^2$, and in any orthonormal tangent
frame with coordinates $\mathbf v=\Log_\mu X$ and
$\Sigma=\mathbb E_P\mathbf v\mathbf v^\top$ one has $D=\tr\Sigma$. The same
statements hold on the real slice with its one-dimensional metric.
\end{theorem}

\begin{proof}
\emph{Finiteness and continuity.} For $p$ in a compact set $K$ and every $x$,
$d(p,x)^2\le 2d(o,x)^2+2\sup_{p\in K}d(o,p)^2$ by the triangle inequality, and
the right side is $P$-integrable by Eq.~\eqref{eq:secondmoment}. Dominated
convergence and continuity of $d$ give continuity of $F$ on $K$.

\emph{Coercivity.} The $L^2$ triangle inequality gives
$(\mathbb E\,d(p,X)^2)^{1/2}\ge d(p,o)-(\mathbb E\,d(o,X)^2)^{1/2}$, hence
$F(p)\to\infty$ as $d(p,o)\to\infty$. A complete finite-dimensional Riemannian
manifold has compact closed bounded balls (Hopf--Rinow, available here because
Theorem~\ref{thm:hadamard} proves completeness), so a minimum exists.

\emph{Strict convexity.} Let $p_t$ be the geodesic from $p_0$ to $p_1$. The
squared-distance inequality on a manifold with $K\le0$ states
\begin{equation}
 \begin{aligned}
 F(p_t)\le{}&(1-t)F(p_0)+tF(p_1)\\
 &-\tfrac12t(1-t)\,d(p_0,p_1)^2 .
 \end{aligned}
 \label{eq:strongconvex}
\end{equation}
It follows from the second variation of $\tfrac12d(\cdot,x)^2$: the Hessian is at
least the metric when $K\le0$, including at $x=p$ by smoothness of the squared
distance there; integrating twice along $p_t$ and taking expectation gives
Eq.~\eqref{eq:strongconvex}. The negative quadratic term forces strict convexity,
hence a unique minimizer.

\emph{First variation.} For $p\ne x$ the gradient of $\tfrac12d(p,x)^2$ is
$-\Log_p x$, whose norm equals $d(p,x)\le d(p,o)+d(o,x)$ and is locally
integrable by Eq.~\eqref{eq:secondmoment}. Dominated differentiation therefore
gives $\nabla F(p)=-\mathbb E_P\Log_p X$, and stationarity at $\mu$ is
Eq.~\eqref{eq:firstorder}.

\emph{Dispersion.} Orthonormal coordinates preserve the norm, so
$D=\mathbb E\|\mathbf v\|^2=\mathbb E\tr(\mathbf v\mathbf v^\top)=\tr\Sigma$.
The real slice is the totally geodesic submanifold $\theta\in\{0,\pi\}$ of
Theorem~\ref{thm:hadamard}, and the same argument applies verbatim.
\end{proof}

A diameter is a closed totally geodesic convex subset: its complete geodesic is
minimizing between every two of its points. A law supported on the real axis is
fixed by conjugation of the disk, so uniqueness in Theorem~\ref{thm:center}
forces its centre to be fixed by that reflection and hence to lie on the
diameter. For such a law the theorem reduces to
\begin{equation}
 \psi(\mu)=\mathbb E_P\psi(X),\qquad D=\Var_P\psi(X).
 \label{eq:realmoments}
\end{equation}

\begin{definition}[Signed isolated-deviation score]
\label{def:sid}
For a reference law supported on the declared real diameter and an observation
$x$ on that diameter, assume $0<D<\infty$. The signed score is
\begin{equation}
 Z_{\mathrm{SID}}(x)=\frac{\psi(\rho(x))-\mathbb E_P\psi(X)}{\sqrt D},
 \label{eq:realZ}
\end{equation}
where the orientation is the declared increasing direction of the real
connection.
\end{definition}

Changing the declared orientation replaces $Z$ by $-Z$ and leaves $Z^2$
unchanged. The sign is information: it distinguishes a connection displaced
toward larger real correlation from one displaced toward smaller real correlation, and no
function of $Z^2$ can recover it.

Both $\psi(r)$ and Fisher's $\operatorname{atanh}(r)$ increase strictly on
the real diameter. With a fixed orientation and positive reference standard
deviations, either standardized coordinate therefore gives identical
one-sided ranks, including ties. Their numerical displacements differ.
For two-sided deviations the ordering also depends on the fitted centre
and the relative stretching on either side. A complex connection has two
tangent coordinates, so this one-dimensional ordering argument does not
apply to its joint score.

\begin{proposition}[Joint score, whitening and rank]
\label{prop:whitening}
Let $\mathbf v=\Log_\mu X$ and $\Sigma=\mathbb E\mathbf v\mathbf v^\top$. If
$\Sigma\succ0$ and $\mathbf W=\Sigma^{-1/2}\mathbf v$, then
$\mathbb E\mathbf W=0$, $\Cov(\mathbf W)=I_2$ and, with
$Q=\|\mathbf W\|^2$, $\mathbb E Q=2$. Under an orthogonal change of tangent frame
$\mathbf v\mapsto O\mathbf v$ and $\Sigma\mapsto O\Sigma O^\top$, so $Q$ is
unchanged although the components of $\mathbf v$ rotate. If $\Sigma$ has rank
$k\in\{0,1,2\}$, then $\mathbf v\in\operatorname{range}(\Sigma)$ almost surely and
\begin{equation}
 \mathbb E\left[\mathbf v^\top\Sigma^{+}\mathbf v\right]=k .
 \label{eq:rankscore}
\end{equation}
Equation~\eqref{eq:rankscore} is a support score, not full-dimensional
whitening.
\end{proposition}

\begin{proof}
Centring gives $\mathbb E\mathbf W=\Sigma^{-1/2}\mathbb E\mathbf v=0$ by
Eq.~\eqref{eq:firstorder}, and
$\Cov(\mathbf W)=\Sigma^{-1/2}\Sigma\Sigma^{-1/2}=I_2$; taking traces gives
$\mathbb E Q=2$. For the frame change, $\Sigma^{-1/2}$ transforms to
$O\Sigma^{-1/2}O^\top$, so $\|\mathbf W\|^2$ is preserved. If
$\Sigma\mathbf a=0$, then $\mathbb E(\mathbf a^\top\mathbf v)^2=\mathbf a^\top
\Sigma\mathbf a=0$, so $\mathbf a^\top\mathbf v=0$ almost surely; a finite basis
of the nullspace gives the support statement. Spectral decomposition of
$\Sigma^+$ yields $\mathbb E[\mathbf v^\top\Sigma^+\mathbf v]=\tr(\Sigma^+\Sigma)=k$.
\end{proof}

A law supported on a complete geodesic through $\mu$ with positive dispersion has
rank one; a point mass has rank zero. Conversely, rank-one covariance implies
that the logarithms lie on a line, so the support lies on that geodesic. Rank
zero is a point mass and admits no standardized score; infinite second moments
invalidate both the centre-dispersion construction and the covariance claim. For
example, place $X=\Exp_o(T\mathbf e)$ with a real Cauchy variable $T$; then
$d(o,X)=|T|$ and Eq.~\eqref{eq:secondmoment} fails.

\begin{remark}[Whitening is a moment identity, not a law]
\label{rem:notgaussian}
Let $\mathbf v$ be uniform on the four vectors $\pm\sqrt2\mathbf e_1,\pm\sqrt2
\mathbf e_2$ at the origin and put $X=\Exp_o\mathbf v$. Then the mean logarithm is
zero and the covariance is $I_2$, so $Q=2$ almost surely, yet the law of $Q$ is a
point mass, the coordinates are dependent, and $Q$ is not chi-square. A
$\chi^2_2$ law follows only from the additional hypothesis
$\mathbf v\sim N(0,\Sigma)$; it is not supplied by the geometry, by finite
moments, or by a central limit theorem for estimated centres.
\end{remark}

\begin{figure*}[t]%
  \centering
  \includegraphics[width=\textwidth]{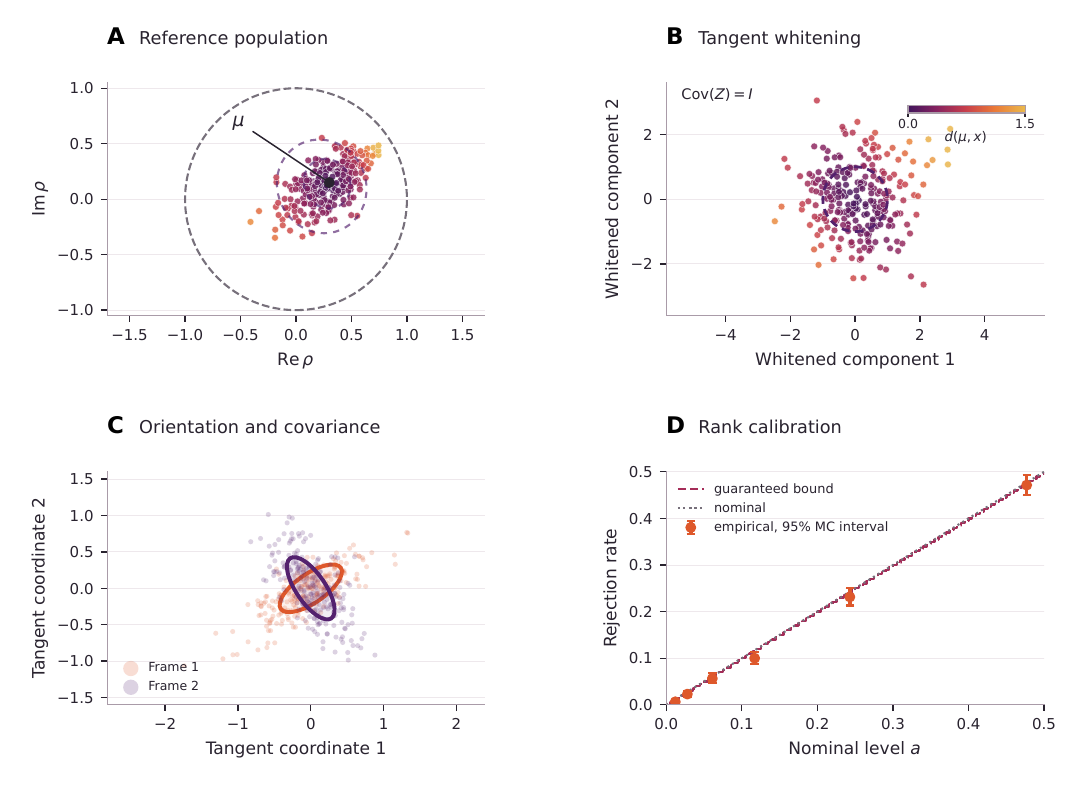}%
  \caption{\textbf{From a reference population to calibrated intrinsic scores.} (A) A Gaussian tangent law mapped to the disk, with geodesic-distance contours about its centre. (B) The same observations after population whitening; the dashed unit circle illustrates identity covariance. In A and B, color tracks each observation's intrinsic displacement. (C) Rotated covariance ellipses with the same intrinsic spectrum. (D) Split calibration of a frozen fitted score with ties. Error bars are pointwise 95\% binomial Monte Carlo intervals; the training fit is held fixed. Computational details are in the Supplementary Methods.}\label{fig:figure_3.pdf}%
  \end{figure*}

Empirical eigensupport is not proof of population support. A rank decision made
from estimated eigenvalues is a decision about the estimated covariance, and the
calibration theory of Section~\ref{sec:calibration} is stated for exactly that
frozen object. Figure~\ref{fig:figure_3.pdf} shows a declared reference law on the
disk with its intrinsic centre, the same observations in the tangent space at that
centre after population whitening, two laws that share an intrinsic covariance spectrum
and differ by a rotation, and the resulting rank calibration.

\section{Ambient and quotient comparisons}
\label{sec:comparison}

\begin{figure*}[t]%
  \centering
  \includegraphics[width=\textwidth]{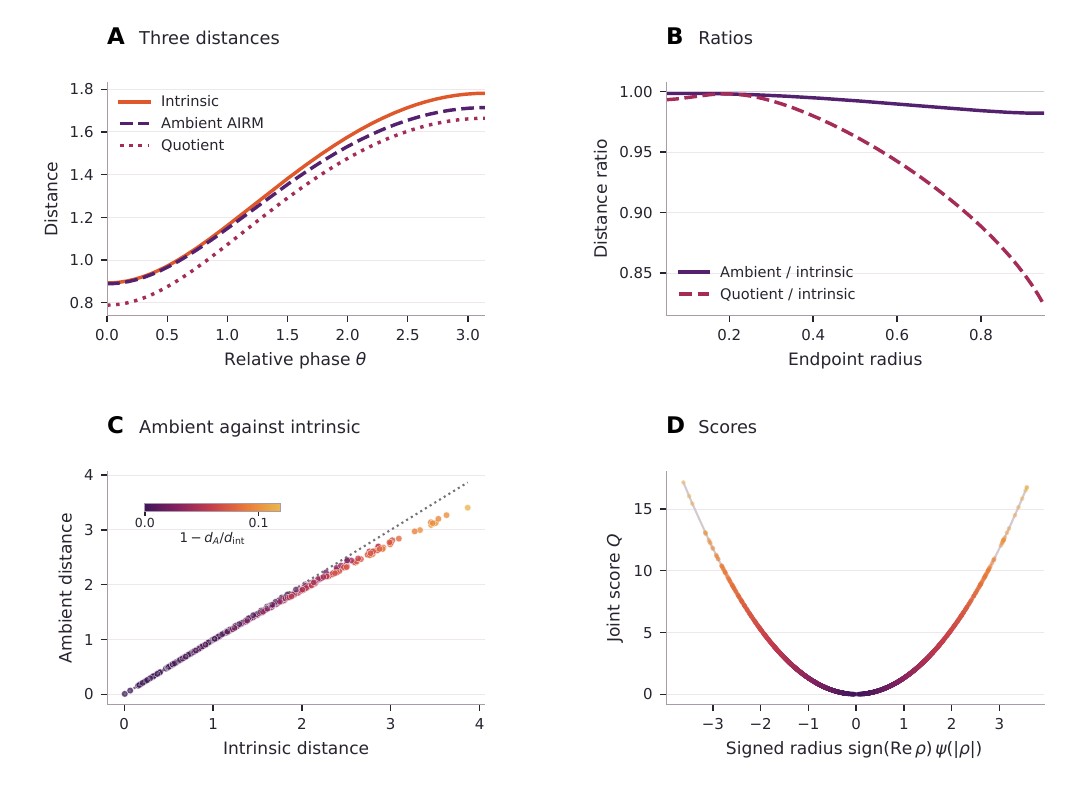}%
  \caption{\textbf{The three distances retain different information.} (A) Distances between identical endpoints as their relative phase varies. (B) Ambient and quotient distances relative to intrinsic distance. (C) Ambient versus intrinsic distance for sampled pairs; color measures the relative distance deficit. (D) The joint score is even in the signed radial coordinate and therefore loses displacement direction. The distance inequalities follow from the geometric results in the text.}\label{fig:figure_4.pdf}%
  \end{figure*}

Three constructions are routinely applied to matrix-valued observations, and they
answer different questions. The ambient affine-invariant geometry lets every
entry vary. A diagonal-scale quotient removes the scale directions. The present
construction holds the complement fixed. This section computes all three on the
same normalized endpoints, so that the comparison is exact rather than
qualitative. Figure~\ref{fig:figure_4.pdf} compares them as the relative phase
varies and separates the signed one-dimensional score from its rotationally
invariant joint counterpart.

The ambient affine-invariant distance and logarithm are
\begin{align}
 d_A(P,Q)&=\bigl\|\log(P^{-1/2}QP^{-1/2})\bigr\|_F,\\
 \Log^A_PQ&=P^{1/2}\log(P^{-1/2}QP^{-1/2})P^{1/2},
 \label{eq:ambientlog}
\end{align}
with the real Hermitian Frobenius norm
$\|H\|_F^2=\operatorname{Re}\tr(H^2)$.

\begin{theorem}[Exact ambient comparison on one slice]
\label{thm:ambientcomparison}
For $p,q\in\D$ write $\Delta_p=1-|p|^2$, $\Delta_q=1-|q|^2$ and
\begin{equation}
 \begin{aligned}
 \ell&=\tfrac12\log\frac{\Delta_q}{\Delta_p},\qquad
 h=\frac{1-\operatorname{Re}(\bar pq)}{\sqrt{\Delta_p\Delta_q}},\\
 k&=\acosh h .
 \end{aligned}
 \label{eq:ambienthk}
\end{equation}
Then $h\ge1$, and
\begin{equation}
 d_A\!\left(A(p),A(q)\right)^2=2\ell^2+2k^2 .
 \label{eq:ambientexact}
\end{equation}
For distinct endpoints the intrinsic distance of
Theorem~\ref{thm:hadamard} is strictly larger than this ambient distance.
\end{theorem}

\begin{proof}
By Theorem~\ref{thm:reduction} only the two-by-two blocks matter. The
eigenvalues $\lambda_1,\lambda_2$ of the pencil $R(p),R(q)$ satisfy
\[
 \lambda_1\lambda_2=\frac{\det R(q)}{\det R(p)}=\frac{\Delta_q}{\Delta_p},
 \qquad
 \lambda_1+\lambda_2=\frac{2\left(1-\operatorname{Re}(\bar pq)\right)}{\Delta_p},
\]
and they are positive because the pencil is congruent to an HPD matrix. Put
$\lambda_{1,2}=\sqrt{\Delta_q/\Delta_p}\,e^{\pm\log\mu}$. The product condition is
satisfied identically and the sum condition becomes
$\mu+\mu^{-1}=2h$. The arithmetic--geometric mean inequality gives $h\ge1$, and
$\mu+\mu^{-1}=2h$ is solved by $\mu=e^{\pm k}$. Therefore
$\log\lambda_{1,2}=\ell\pm k$ and
$\sum_i(\log\lambda_i)^2=2\ell^2+2k^2$, which is
Eq.~\eqref{eq:ambientexact}.

Every slice path is an ambient path, so $d_A\le d$. Suppose equality held for
distinct endpoints. Then the unique minimizing slice geodesic would be an
ambient minimizing geodesic. The ambient geodesic equation is
$A''=A'A^{-1}A'$, obtained by differentiating the exponential in
Eq.~\eqref{eq:ambientlog}. At any point where $z'\neq0$, the first diagonal entry
satisfies
\[
 e_1^*A''e_1=e_1^*A'A^{-1}A'e_1>0,
\]
because $A'e_1\neq0$ and $A^{-1}\succ0$. A slice curve has constant first
diagonal entry, hence $e_1^*A''e_1=0$ identically. A nonconstant slice geodesic has constant positive speed. Since $z$ is its
only varying coordinate, $z'\ne0$ at every parameter value, yielding the
contradiction.
\end{proof}

For example, the equal-weight ambient mean of $R(r)$ and $R(-r)$, $0<r<1$, is
$\sqrt{1-r^2}\,I_2$ because the matrices commute, whereas the intrinsic slice
mean is $R(0)=I_2$. The ambient mean has changed both diagonal entries; it has
left the slice. At the slice reference point
\begin{equation}
 \log R(\rho)=\tfrac12\log(1-r^2)I_2+
 \frac{\operatorname{atanh}r}{r}
 \begin{pmatrix}0&\rho\\ \bar\rho&0\end{pmatrix},
 \label{eq:ambientzero}
\end{equation}
with the ratio equal to one at the origin. The edge component is the ambient
radial vector of length $\sqrt2\operatorname{atanh}r$, while the intrinsic slice
logarithm has length $\psi(r)$. Their first-order agreement follows from
$\psi(r)=\sqrt2r+O(r^3)$ and does not identify their finite-displacement scores
or their covariances.

\subsection{The two-variable diagonal-scale quotient}

Complex correlation quotient geometry was formulated by David and
Gu~\cite{G01}; the real inverse-hyperbolic-tangent distance and the associated
anomaly scores are established in~\cite{G02,G03}, and curvature restrictions on
the full quotient-affine manifold in~\cite{G04}. Here we calculate the
two-variable comparator in the same normalization as Eq.~\eqref{eq:airm}, so that
the three geometries can be compared directly.

\begin{proposition}[Exact complex two-variable quotient]
\label{prop:quotient}
Quotienting $\HPD(2)$ by positive real diagonal congruence gives, on
$\rho=r\mathrm e^{\mathrm i\theta}$,
\begin{equation}
 \dd s_Q^2=\frac{2\,\dd r^2}{(1-r^2)^2}+\frac{2r^2\,\dd\theta^2}{1-r^2}.
 \label{eq:quotientmetric}
\end{equation}
Its Gaussian curvature is the constant $-1/2$, and its distance is
$d_Q(p,q)=\sqrt2\,k$ with $k$ from Eq.~\eqref{eq:ambienthk}. Consequently
\begin{equation}
 d_A^2=d_Q^2+2\ell^2
 \label{eq:decomposition}
\end{equation}
on the normalized representatives, and the real restriction is
$\sqrt2\,|\operatorname{atanh}p-\operatorname{atanh}q|$.
\end{proposition}

\begin{proof}
At $\rho=r$ the diagonal-scale directions are spanned by $R$ and
$J=\diag(1,-1)$. A radial edge derivative $U$ satisfies
$g(U,R)=-2r/(1-r^2)$, $g(R,R)=2$ and $g(U,J)=0$; the phase derivative is
orthogonal to both. The span is orthogonal in this basis because
$g(R,J)=\tr(R^{-1}J)=0$. Removing the radial projection of the vertical
directions subtracts $2r^2/(1-r^2)^2$ from the radial coefficient of
Eq.~\eqref{eq:polar} and leaves the angular coefficient unchanged. Rephasing
transports the calculation to every $\rho$, which gives
Eq.~\eqref{eq:quotientmetric}.

Put $u=\sqrt2\operatorname{atanh}r$; the warped-product profile is
$\sqrt2\sinh(u/\sqrt2)$, whose second derivative divided by itself is $1/2$, so
the curvature is $-1/2$. Equivalently, the map
\begin{equation}
 X(\rho)=\frac{(1,x,y)}{\sqrt{1-r^2}}
 \label{eq:hyperboloid}
\end{equation}
pulls the metric $2(-\dd X_0^2+\dd X_1^2+\dd X_2^2)$ on the future unit
hyperboloid back to Eq.~\eqref{eq:quotientmetric}. The Lorentz inner product of
$X(p)$ and $X(q)$ is $-h$. The intersection of the hyperboloid with their
two-plane is a geodesic of length $\sqrt2k$, parameterized by
\[
 \sinh((1-t)k)\frac{X(p)}{\sinh k}+\sinh(tk)\frac{X(q)}{\sinh k},\qquad 0\le t\le1,
\]
and completeness with negative curvature makes it the unique shortest path.
Equation~\eqref{eq:decomposition} is then Eq.~\eqref{eq:ambientexact} with the
factor $2\ell^2$ split off. Restricting to a diameter gives the stated real
formula.
\end{proof}

This calculation concerns the two-variable Schur block, not the quotient of a
full $n$-variable correlation matrix, and confers neither constant curvature nor
uniqueness on the latter; the general limitations are established
in~\cite{G04,G05}. For implementation the hyperboloid logarithm is
$V=k[X(q)-hX(p)]/\sinh k$ with the continuous zero-distance limit, pulled back by
$\dd\rho=(\dd X_1,\dd X_2)/X_0-(X_1,X_2)\dd X_0/X_0^2$.

\subsection{Flat coordinate comparators}

The log-Euclidean comparator uses the isometric Hermitian vectorization of
$\log A$ with Euclidean centring and covariance~\cite{G20,G21}. It extends to
complex HPD because the principal Hermitian logarithm is a smooth bijection onto
the space of Hermitian matrices, with inverse the matrix exponential. The
log-Cholesky construction in the real case is due to Lin~\cite{G16}; the complex
extension used here is the following pullback, which we derive rather than
attribute to~\cite{G16}.

Let $A=LL^*$ be the Cholesky factorization with positive real diagonal and define
$c(A)$ to contain $\log L_{jj}$ together with the unweighted real and imaginary
parts of $L_{ij}$ for $i>j$. These coordinates are a diffeomorphism onto
$\R^{n^2}$: exponentiating the diagonal and filling the strict lower triangle
reconstructs $L$ and then $A$, and the positive pivots in the recursive Cholesky
formula make both the map and its inverse smooth. Euclidean distance,
straight-line coordinate geodesics, coordinate averaging and covariance follow by
this explicit pullback. There is no $\sqrt2$ weight here, because the coordinate
metric is defined by this choice rather than by Hermitian Frobenius
vectorization. On a fixed slice $A_{11}=a$ is constant, so $\log L_{11}
=\tfrac12\log a$ is removed before covariance fitting.

Log-Cholesky depends on the ordering of the variables, and arbitrary rescalings
change its strict-triangle coordinates differently from its log diagonal. For
dimensional data a fixed declared unit scale is required before taking logarithms
or choosing this coordinate metric, and no universal invariance of its whitened
score is presumed.

\section{Invariances of the declared slice}
\label{sec:invariance}

The slice carries a group of transformations that preserve its normalized
geometry exactly. Knowing that group, and knowing what it does not contain, is
what separates a statement about a geometric object from a statement about a
particular numerical matrix.

\begin{proposition}[Slice-preserving congruences]
\label{prop:invariances}
Let $T=\diag(d_1,d_2,W)$ with $d_1,d_2\in\C\setminus\{0\}$ and $W$ invertible
on the complement. Then $T$ maps one fixed-complement slice onto another, and
\begin{equation}
 \begin{gathered}
 \alpha'=|d_1|^2\alpha,\quad \beta'=|d_2|^2\beta,\quad
 \gamma'=d_1\bar d_2\gamma,\\
 \rho'=e^{\mathrm i\varphi}\rho,\quad
 e^{\mathrm i\varphi}=\frac{d_1\bar d_2}{|d_1d_2|}.
 \end{gathered}
 \label{eq:sliceaction}
\end{equation}
The induced action on the disk is a rotation. It is an isometry of
Eq.~\eqref{eq:cartmetric}. Swapping the two selected labels conjugates $\rho$; a
permutation that transports the selected pair and the complement only relabels
the construction. In the real case, positive rescaling leaves $\rho$ unchanged
while real rescaling with $d_1d_2<0$ reverses the signed coordinate.
\end{proposition}

\begin{proof}
Under $T$ the blocks become $C'=WCW^*$, $E'=\diag(d_1,d_2)EW^*$ and
$a'=|d_1|^2a$, $b'=|d_2|^2b$. Then
$H'=E'C'^{-1}E'^*=\diag(d_1,d_2)EC^{-1}E^*\diag(\bar d_1,\bar d_2)$, because
$W$ and $W^{-1}$ cancel in $E'C'^{-1}E'^*$. Reading off the entries gives the
first three identities of Eq.~\eqref{eq:sliceaction}, and
\[
 \rho'=\frac{d_1\bar d_2(z-\gamma)}
 {\sqrt{|d_1|^2\alpha\,|d_2|^2\beta}}
 =\frac{d_1\bar d_2}{|d_1d_2|}\,\rho .
\]
A rotation of the disk preserves $r=|\rho|$, and by Eq.~\eqref{eq:polar} it
preserves $G$. Swapping the two selected labels exchanges $a\leftrightarrow b$
and sends $z\mapsto\bar z$, hence $\rho\mapsto\bar\rho$. For real factors, $e^{\mathrm i\varphi}=\operatorname{sign}(d_1d_2)$. The signed coordinate reverses exactly when $d_1d_2<0$.
\end{proof}

If the reference law is pushed forward by the same action, its centre,
logarithms and covariance push forward under the differential, so the dispersion
and the scalar $Q$ of Proposition~\ref{prop:whitening} are invariant while signed
or frame components obey their rotation or reflection rules. Using a different
reference law without pushing the original one forward is a different
statistical problem and carries no such invariance.

\begin{proposition}[Fixed complementary regressors]
\label{prop:regressors}
Fix $R_0$ of compatible size and invertible $W$, and let
$T=\begin{psmallmatrix}D&R_0\\0&W\end{psmallmatrix}$ with
$D=\diag(d_1,d_2)$ (or a selected-label permutation followed by that diagonal).
Then the transformed cross block is $(DE+R_0C)W^*$ and remains fixed across the
slice. For diagonal $D$ the normalized connection transforms as in
Eq.~\eqref{eq:sliceaction}; an additional swap of the selected labels
composes that rotation with complex conjugation.
\end{proposition}

\begin{proof}
The cross block of $TAT^*$ in the selected rows is $DE+R_0C$ times $W^*$, which
does not depend on $z$. Expanding the Schur complement of the transformed matrix
and cancelling the three terms introduced by $R_0$ leaves
$S'=DSD^*$. For diagonal $D$, normalizing by
$\diag(|d_1|\sqrt\alpha,|d_2|\sqrt\beta)$ gives Eq.~\eqref{eq:sliceaction}.
A swap exchanges the two diagonal entries and conjugates the normalized
off-diagonal entry, as in Proposition~\ref{prop:invariances}.
\end{proof}

Proposition~\ref{prop:regressors} describes a sufficient family, not an
exhaustive classification of slice-preserving transformations.

\begin{remark}[What ambient invariance does not transfer]
\label{rem:noambient}
Generic mixing and inversion preserve the ambient affine-invariant distance but
do not preserve the isolated-entry constraint. The real Hadamard rotation
$T=2^{-1/2}\begin{psmallmatrix}1&1\\1&-1\end{psmallmatrix}$ sends $R(t)$ to
$\diag(1+t,1-t)$, so a varying edge has become a varying diagonal. Likewise
\[
 R(t)^{-1}=(1-t^2)^{-1}\begin{pmatrix}1&-t\\-t&1\end{pmatrix}
\]
has varying diagonal entries: its first diagonal derivative vanishes at $t=0$ but
its second derivative does not. A pointwise inverse normalization that happens to
reproduce $-t$ is therefore not a constant slice-preserving matrix action.
\end{remark}

These examples prohibit unrestricted transfer of ambient invariance to a labelled
connection score. An isolated tensor connection retains its physical frame, and
covariance or coherence conditioning by itself establishes no directed causal
interpretation. The conditional spectral quantities built on the same Schur block
carry a directional reading only when a directional model is supplied on top of
them, as in partial-coherence tests for Granger
causality~\cite{G39,G40}.

The induced geometry supplies a fixed-complement setting for established tools from correlation geometry, Fr\'echet averaging, tangent whitening and exchangeable rank calibration~\cite{G01,G03,G04,S03,S10,S14}.

\section{Geodesics, logarithms and parallel transport}
\label{sec:geodesics}

Theorem~\ref{thm:hadamard} guarantees that every pair of points is joined by one
minimizing geodesic. This section turns that existence statement into a
computable formula: the geodesic problem reduces to one scalar root on two
exactly integrable legs, and the logarithm is recovered from the same data.
Figure~\ref{fig:figure_5.pdf} shows three endpoint configurations from one base point, the turning branch at a wider separation, the three distances on identical endpoints and the scalar reduction against independent shooting.

\subsection{Christoffel symbols and the geodesic equation}

Write $x=(x_1,x_2)$, $q=|x|^2$, and let $\delta$ denote the Kronecker symbol.

\begin{lemma}[Levi-Civita symbols]
\label{lem:christoffel}
For the metric~\eqref{eq:cartmetric} the Christoffel symbols of the Levi-Civita
connection are
\begin{equation}
 \Gamma^i_{jk}=\frac{x_j\delta_{ik}+x_k\delta_{ij}}{1-q}
 +\frac{x_i\delta_{jk}}{1+q}.
 \label{eq:christoffel}
\end{equation}
The geodesic equation is therefore
\begin{equation}
 \ddot x=-\frac{2(x\cdot\dot x)\,\dot x}{1-q}-\frac{|\dot x|^2\,x}{1+q}.
 \label{eq:cartode}
\end{equation}
\end{lemma}

\begin{proof}
The Levi-Civita connection is the unique torsion-free connection compatible with
$G$, so it suffices to verify those two properties for
Eq.~\eqref{eq:christoffel} and to lower an index. Symmetry in $j,k$ is immediate
from Eq.~\eqref{eq:christoffel}, which is torsion-freeness.

For compatibility, put $\lambda=2/(1-q)$ and $\eta=4/(1-q)^2$, so that
$G_{ij}=\lambda\delta_{ij}+\eta x_ix_j$. Then
\[
 \partial_m\lambda=\frac{4x_m}{(1-q)^2},\qquad
 \partial_m\eta=\frac{16x_m}{(1-q)^3},
\]
and
\[
 \partial_kG_{ij}=\frac{4x_k\delta_{ij}}{(1-q)^2}
 +\eta\left(\delta_{ik}x_j+x_i\delta_{jk}\right)
 +\frac{16x_ix_jx_k}{(1-q)^3}.
\]
On the other hand
$G_{il}\Gamma^l_{jk}=\lambda\Gamma^i{}_{jk}+\eta x_i x_l\Gamma^l{}_{jk}$, and
\[
 x_l\Gamma^l{}_{jk}=\frac{2x_jx_k}{1-q}+\frac{q\,\delta_{jk}}{1+q}.
\]
Adding the corresponding expression with $i$ and $j$ interchanged, and using
$\eta=\lambda^2$ together with $\partial_k\lambda=4x_k/(1-q)^2$ and
$\partial_k\eta=16x_k/(1-q)^3$, reproduces exactly $\partial_kG_{ij}$: the
$\delta_{ik}$ coefficient becomes
$2x_j/(1-q)^2+2x_j/[(1-q)(1+q)]+4qx_j/[(1-q)^2(1+q)]=4x_j/(1-q)^2$, the
$\delta_{ij}$ coefficient becomes $4x_k/(1-q)^2$, and the cubic term becomes
$4\eta x_ix_jx_k/(1-q)=16x_ix_jx_k/(1-q)^3$. Hence
$\partial_kG_{ij}=G_{il}\Gamma^l{}_{jk}+G_{jl}\Gamma^l{}_{ik}$, which is metric
compatibility. By the fundamental theorem of Riemannian geometry this is the
Levi-Civita connection of $G$.

Contracting Eq.~\eqref{eq:christoffel} with $\dot x^j\dot x^k$ gives
\[
 \Gamma^i{}_{jk}\dot x^j\dot x^k
 =\frac{2(x\cdot\dot x)\dot x^i}{1-q}+\frac{|\dot x|^2x^i}{1+q},
\]
and $\ddot x^i=-\Gamma^i{}_{jk}\dot x^j\dot x^k$ is
Eq.~\eqref{eq:cartode}.
\end{proof}

The Cartesian form is regular everywhere in the open disk; the apparent
singularity in the polar symbols at $r=0$ is the coordinate artefact already
noted after Theorem~\ref{thm:metric}. For reference,
\begin{equation}
 \begin{gathered}
 \Gamma^r_{rr}=\frac{r}{1+r^2}+\frac{2r}{1-r^2},\quad
 \Gamma^r_{\theta\theta}=-\frac{r}{1+r^2},\\
 \Gamma^\theta_{r\theta}=\frac{1}{r(1-r^2)}.
 \end{gathered}
 \label{eq:polarchristoffel}
\end{equation}

\subsection{First integrals and the scalar reduction}

Because $G$ is invariant under rotations, the geodesic flow has two first
integrals. Parametrize a geodesic by arc length, so $g(\dot x,\dot x)=1$, and
write $s=\psi(r)$ and $w=f(s)=\sqrt2\,r/\sqrt{1-r^2}$.

\begin{lemma}[Clairaut constant and energy]
\label{lem:clairaut}
Along a unit-speed geodesic the quantity
\begin{equation}
 c:=f(s)^2\,\dot\theta=w^2\dot\theta
 \label{eq:clairaut}
\end{equation}
is constant, and
\begin{equation}
 \dot s^{\,2}+\frac{c^2}{w^2}=1 .
 \label{eq:energy}
\end{equation}
In particular $w\ge|c|$ wherever the geodesic is defined, with equality exactly
at a radial turning point, and a non-radial geodesic has at most one turning
point. The constant $c$ is signed, carrying the direction of rotation.
\end{lemma}

\begin{proof}
In polar coordinates the Lagrangian of the metric~\eqref{eq:polar} is
$\mathcal L=\tfrac12\bigl(a(r)\dot r^2+b(r)\dot\theta^2\bigr)$ and does not depend
on $\theta$, so $p_\theta=b(r)\dot\theta$ is conserved. Since
$b(r)=2r^2/(1-r^2)=f(s)^2$, this is Eq.~\eqref{eq:clairaut}. Unit speed gives
$a\dot r^2+b\dot\theta^2=1$, and $a=\psi'^2$ together with $\dot s=\psi'\dot r$
turns this into Eq.~\eqref{eq:energy}. Equation~\eqref{eq:energy} forces
$w^2\ge c^2$, and equality forces $\dot s=0$. The second derivative of $s$ at a
turning point is $\ddot s=f f_s\dot\theta^2>0$ whenever $c\ne0$, so the radius has a
strict minimum there and cannot turn twice.
\end{proof}

Set $h(w)=\dd s/\dd w$, the inverse of $\dd w/\dd s$. From
$w=\sqrt2 r/\sqrt{1-r^2}$ one computes $r(w)=w/\sqrt{w^2+2}$ and
\begin{equation}
 h(w)=\sqrt{1-r(w)^4}=\frac{2\sqrt{1+w^2}}{w^2+2}.
 \label{eq:hw}
\end{equation}
The function $h$ is positive, satisfies $h(0)=1$, and is strictly decreasing for
$w>0$.

\begin{lemma}[Closed-form geodesic legs]
\label{lem:legs}
For $0<c\le W$ define
\begin{align}
 A(c,W)&=\int_c^W\frac{c\,h(w)}{w\sqrt{w^2-c^2}}\,\dd w\\
 &=\int_0^{\arccos(c/W)}h(c\sec t)\,\dd t,
 \label{eq:Adef}\\
 L(c,W)&=\int_c^W\frac{w\,h(w)}{\sqrt{w^2-c^2}}\,\dd w\\
 &=\int_0^{\acosh(W/c)}c\cosh u\;h(c\cosh u)\,\dd u .
 \label{eq:Ldef}
\end{align}
With
\begin{equation}
 \begin{gathered}
 D=\sqrt{W^2-c^2},\quad a=\sqrt{1+c^2},\\
 b=\sqrt{2+c^2},\quad v=\frac{D}{\sqrt{1+W^2}},
 \end{gathered}
 \label{eq:legsymbols}
\end{equation}
both legs are elementary:
\begin{align}
 A(c,W)&=\arctan\frac{v}{c}+\frac{c}{b}\operatorname{atanh}\frac{v}{b},
 \label{eq:Aclosed}\\
 L(c,W)&=2\left(\operatorname{asinh}\frac{D}{a}
 -\frac1b\operatorname{atanh}\frac{v}{b}\right).
 \label{eq:Lclosed}
\end{align}
Moreover $A(0,W)=\pi/2$, $L(0,W)=\psi\bigl(r(W)\bigr)$, $A(c,c)=L(c,c)=0$, and
both legs are continuous on $\{0\le c\le W,\ W>0\}$ together with the limits
$c\downarrow0$ at fixed $W>0$. The angular leg has no continuous extension to $(0,0)$: along $c=0$
its value is $\pi/2$, whereas along $c=W>0$ it is zero. The length leg does
extend continuously there with value zero.
\end{lemma}

\begin{proof}
The substitution $w=c\sec t$ in Eq.~\eqref{eq:Adef} gives
$\dd w=w\tan t\,\dd t$ and $\sqrt{w^2-c^2}=c\tan t$, hence
$\dd w/(w\sqrt{w^2-c^2})=\dd t/c$ and the second form. The substitution
$w=c\cosh u$ in Eq.~\eqref{eq:Ldef} gives $\dd w/\sqrt{w^2-c^2}=\dd u$ and the
second form there.

For Eq.~\eqref{eq:Aclosed}, differentiate
$F(W)=\arctan(v/c)+(c/b)\operatorname{atanh}(v/b)$ with respect to $W$. Since
$D'=W/D$ and $v'=\bigl(W/D-Wv/\sqrt{1+W^2}\bigr)/\sqrt{1+W^2}$, an
algebraic simplification gives
\begin{equation}
 F'(W)=\frac{c\,h(W)}{W\sqrt{W^2-c^2}},
 \label{eq:Aderiv}
\end{equation}
which is the integrand of Eq.~\eqref{eq:Adef} evaluated at the upper limit, and
$F(c)=0$. Hence $F=A(c,\cdot)$. For Eq.~\eqref{eq:Lclosed}, the same computation
for $G(W)=2\bigl(\operatorname{asinh}(D/a)-(1/b)\operatorname{atanh}(v/b)\bigr)$
gives
\begin{equation}
 G'(W)=\frac{W\,h(W)}{\sqrt{W^2-c^2}},
 \label{eq:Lderiv}
\end{equation}
which is the integrand of Eq.~\eqref{eq:Ldef}, and $G(c)=0$. For fixed $W>0$, extend the angular integrand by zero to $[0,\pi/2]$.
It is bounded by one and converges pointwise almost everywhere to one as
$c\downarrow0$, so dominated convergence gives $A\to\pi/2$.
For the length leg, set $w=\sqrt{c^2+(W^2-c^2)v^2}$ to obtain
$L=\sqrt{W^2-c^2}\int_0^1h(w(v))\dd v$. Dominated convergence gives
$L(0,W)=\int_0^W h(w)\dd w=\psi(r(W))$. Also $0\le L(c,W)\le W$,
which proves continuity of $L$ at $(0,0)$. At $c=W$ both integrals have zero length, and both closed forms
reduce to $\arctan0+0=0$ and $2(\operatorname{asinh}0-0)=0$.
\end{proof}

\begin{proposition}[Existence and uniqueness of the reduction]
\label{prop:root}
Fix non-degenerate endpoints: radii $r_1,r_2>0$ and signed angular separation
$\delta$ with $0<|\delta|<\pi$. Put $W_i=w(r_i)$ and order $W_1\le W_2$, and let
$\delta_*=A(W_1,W_2)$. Then a unique geodesic joins the endpoints, the magnitude of its Clairaut
constant is the unique $c\in(0,W_1]$ solving
\begin{align}
 A(c,W_2)-A(c,W_1)&=|\delta|, && |\delta|\le\delta_*,
 \label{eq:monotone}\\
 A(c,W_2)+A(c,W_1)&=|\delta|, && |\delta|\ge\delta_*.
 \label{eq:turning}
\end{align}
On the first branch the radius varies monotonically and the distance is
$L(c,W_2)-L(c,W_1)$; on the strict second branch $|\delta|>\delta_*$ the geodesic has one
interior radial minimum, and its distance is $L(c,W_2)+L(c,W_1)$. At $|\delta|=\delta_*$ the two branch
equations coincide and both formulas give the same distance.
\end{proposition}

\begin{proof}
For $W_1<W_2$ the monotone angle is
\[
 M(c)=\int_{W_1}^{W_2}\frac{c\,h(w)}{w\sqrt{w^2-c^2}}\,\dd w.
\]
It is strictly increasing for $0<c<W_1$, since the derivative of its
integrand with respect to $c$ is $h(w)w/(w^2-c^2)^{3/2}>0$.
Continuity at the endpoints gives $M(0)=0$ and $M(W_1)=\delta_*$.
If $W_1=W_2$, then $M\equiv0$ and $\delta_*=0$, so no nonzero angle
uses the monotone branch.

For fixed $W$, the transformed angular integral in Eq.~\eqref{eq:Adef}
is strictly decreasing in $c$: its upper limit decreases, and $h(c\sec t)$
decreases pointwise on the common interval. Thus
$T(c)=A(c,W_2)+A(c,W_1)$ decreases strictly from $\pi$ at $c=0$ to
$\delta_*$ at $c=W_1$. These ranges prove existence and uniqueness of
the stated scalar root and identify the branch without a separate
monotonicity assumption.

Conversely, prescribe the unit tangent at the smaller-radius endpoint
with angular component $c/W_1$ and radial component
$\pm\sqrt{1-c^2/W_1^2}$, taking the outward sign for $M$ and inward sign
for $T$. The smooth Cartesian geodesic equation has a unique solution.
Its conserved quantities give the stated angular integrals. In the turning
case the inward part reaches $w=c$ in finite length; the positive radial
acceleration there gives its smooth outward continuation. The root equation
places the other endpoint at precisely its prescribed angle and radius.
Theorem~\ref{thm:hadamard} makes this geodesic minimizing.

The energy identity gives the length element
$\dd t=h(w)w|\dd w|/\sqrt{w^2-c^2}$, and hence the difference or sum of
the length legs. At $|\delta|=\delta_*$, $c=W_1$: the radial minimum is
the smaller-radius endpoint, with no interior turning point, and both
length expressions agree. The sign of $\delta$ fixes the angular direction.
\end{proof}

Degenerate configurations are handled separately: coincident points have
$d=0$; equal phases give $d=|\psi(r_2)-\psi(r_1)|$; opposite phases give
$d=\psi(r_1)+\psi(r_2)$; an endpoint at the origin gives
$d=\psi(r_{\text{other}})$ by Theorem~\ref{thm:hadamard}. Neither wrapping at
$\pm\pi$ nor exchanging the endpoints produces a second segment.

For stable evaluation the monotone branch should be integrated over
$[W_1,W_2]$ rather than formed as a difference of nearly equal legs, and
Eq.~\eqref{eq:Ldef} can be evaluated on the fixed unit interval by the change of
variable $w(v)=\sqrt{c^2+(W^2-c^2)v^2}$:
\begin{equation}
 L(c,W)=\sqrt{W^2-c^2}\int_0^1 h\bigl(w(v)\bigr)\,\dd v ,
 \label{eq:Lstable}
\end{equation}
which also has the correct limit at $c=0$.

\subsection{Logarithms and parallel transport}

At a non-zero base point $p$ use the orthonormal frame
$e_s=\partial_s$, $e_\theta=f^{-1}\partial_\theta$, and let $W_p=w(r_p)$ and $d$
be the distance of Proposition~\ref{prop:root}. Writing $c$ for the solved
Clairaut constant and $\varepsilon=\operatorname{sign}\delta$, the logarithm is
\begin{equation}
 \Log_p(q)=d\left(\sigma\sqrt{1-\frac{c^2}{W_p^2}}\;e_s
 +\varepsilon\frac{c}{W_p}\;e_\theta\right),
 \label{eq:logframe}
\end{equation}
where $\sigma=+1$ if the radius initially increases and $\sigma=-1$ if it
decreases; a turning segment starts inward, and a monotone segment uses
$\operatorname{sign}(r_q-r_p)$. At $c=W_p$ the radial component vanishes. At
$p=0$ the Cartesian frame $2^{-1/2}\partial_{x_1},2^{-1/2}\partial_{x_2}$ is used
instead, with components $\psi(r_q)(\cos\theta_q,\sin\theta_q)$, and coincident
points have zero logarithm.

The Cartesian initial velocity is recovered from Eq.~\eqref{eq:logframe} by
writing $u_s,u_\theta$ for its orthonormal components and using
\[
 v=\frac{u_s}{\sqrt{a(r_p)}}(\cos\theta_p,\sin\theta_p)
 +\frac{u_\theta\sqrt{1-r_p^2}}{\sqrt2}(-\sin\theta_p,\cos\theta_p).
\]
Integrating Eq.~\eqref{eq:cartode} with this initial velocity to time one computes $\Exp_p$.
Algorithm~\ref{alg:distance-log} summarizes the distance and logarithm evaluation.
\begin{table}[!t]
\centering
\refstepcounter{paperalgorithm}\label{alg:distance-log}
{\sffamily\footnotesize
\noindent\textbf{Algorithm~\thepaperalgorithm: Intrinsic distance and logarithm on the connection disk}\par
\vspace{4pt}
\begin{tabular}{@{}p{0.055\columnwidth}@{\hspace{0.025\columnwidth}}>{\raggedright\arraybackslash}p{0.92\columnwidth}@{}}
\toprule
\multicolumn{2}{@{}>{\raggedright\arraybackslash}p{\columnwidth}@{}}{\textbf{Input:} normalized endpoints $p,q\in\D$.
\textbf{Output:} intrinsic distance $d$ and Cartesian logarithm $v=\Log_p(q)$.
For $z\ne0$, let $e(z)=(\operatorname{Re}z,\operatorname{Im}z)/|z|$.}\\[3pt]
\midrule
1 & Reject non-finite endpoints or $|p|\ge1$ or $|q|\ge1$.
If $p=q$, return $(d,v)=(0,0)$. Set $r_p=|p|$ and $r_q=|q|$.\\[3pt]
2 & If $p=0$, return $d=\psi(r_q)$ and $v=d\,e(q)/\sqrt2$.
If $q=0$, return $d=\psi(r_p)$ and $v=-d\,e(p)/\sqrt{a(r_p)}$.\\[3pt]
3 & Set $\delta=\operatorname{Arg}(q/p)\in(-\pi,\pi]$, $\alpha=|\delta|$,
$e_r=e(p)$ and $e_\theta=(-e_{r,2},e_{r,1})$.\\[3pt]
4 & If $\alpha=0$, set $s=\psi(r_q)-\psi(r_p)$ and return
$d=|s|$, $v=s\,e_r/\sqrt{a(r_p)}$.
If $\alpha=\pi$, return $d=\psi(r_p)+\psi(r_q)$ and
$v=-d\,e_r/\sqrt{a(r_p)}$.\\[3pt]
5 & Set $W_p=w(r_p)$, $W_q=w(r_q)$, $W_s=\min(W_p,W_q)$,
$W_l=\max(W_p,W_q)$ and $\alpha_*=A(W_s,W_l)$.\\[3pt]
6 & Select the monotone branch if $\alpha\le\alpha_*$ and the turning branch otherwise.
At equality set $c=W_s$. Otherwise solve on the full bracket $[0,W_s]$:
$A(c,W_l)-A(c,W_s)=\alpha$ for the monotone branch, or
$A(c,W_l)+A(c,W_s)=\alpha$ for the turning branch.\\[3pt]
7 & On the monotone branch set $d=L(c,W_l)-L(c,W_s)$ and
$\sigma=\operatorname{sign}(r_q-r_p)$.
On the turning branch set $d=L(c,W_l)+L(c,W_s)$ and $\sigma=-1$.\\[3pt]
8 & Form $u_s=\sigma d\sqrt{1-c^2/W_p^2}$ and
$u_\theta=\operatorname{sign}(\delta)\,dc/W_p$.
Return $v=u_s e_r/\sqrt{a(r_p)}+
u_\theta\sqrt{1-r_p^2}\,e_\theta/\sqrt2$ together with $d$.\\[3pt]
\midrule
\multicolumn{2}{@{}>{\raggedright\arraybackslash}p{\columnwidth}@{}}{\textbf{Numerical evaluation.}
The branch equalities above specify exact arithmetic. Evaluate the legs and
radial limits stably, and increase precision for close endpoints, small radii
or unresolved branch comparisons. Resolve $c$ and the logarithm components
before rounding to machine precision. Return an unavailable result when a
bracket, finite value or requested tolerance cannot be resolved.
Precision refinement is distinct from the interval enclosures discussed in
Section~\ref{sec:geodesics}.}\\
\bottomrule
\end{tabular}}
\end{table}

For parallel transport, the connection identities away from the origin are
$\nabla e_s=f_s\dd\theta\,e_\theta$ and $\nabla e_\theta=-f_s\dd\theta\,e_s$,
so a transported field $V=ue_s+ve_\theta$ solves
\begin{equation}
 u'=f_s\dot\theta\,v,\qquad v'=-f_s\dot\theta\,u,
 \label{eq:transport}
\end{equation}
a rotation through $-\int f_s\dd\theta$. At the origin the nonsingular Cartesian
equation $\dot V^i+\Gamma^i{}_{jk}\dot x^jV^k=0$ is used. Metric compatibility
proves that transported tangent inner products are preserved along the curve.
The columns of $G^{-1/2}$ supply an orthonormal frame at a single point; they do
not by themselves implement transport between two points separated along a
curve, and no such identification is used here.

\subsection{Numerical error bounds}

The following conditional bounds describe a route to validated numerical
enclosures. The accompanying implementation returns floating-point estimates
with diagnostics; its precision checks do not implement interval certification.

For a continuous non-negative integrand on a finite interval, subdividing the
interval and enclosing the integrand on each cell by outward-rounded interval
arithmetic bounds the integral by the sum of cell widths times the lower and
upper bounds. The transformed legs of Lemma~\ref{lem:legs} have smooth integrands
on their closed intervals for $c>0$, and Eq.~\eqref{eq:Lstable} remains
continuous at $c=0$, so refinement gives mathematical convergence with directed
arithmetic. Branch decisions require disjoint certified angle intervals;
overlapping comparisons must be refined or reported as unresolved.

A complementary a-posteriori route uses a validated enclosure of the Cartesian
flow. If $\hat q=\Exp_p(\hat v)$ is enclosed and $d(\hat q,q)\le\Delta$ is
certified, then
\begin{equation}
 \|\hat v-\Log_p(q)\|_p\le\Delta,
 \label{eq:logcert}
\end{equation}
because on a Hadamard manifold the exponential expands distances and its inverse
is one-Lipschitz. If both the enclosure and $q$ lie in a Euclidean disk of radius
$R<1$ with Euclidean separation at most $e$, the straight segment gives
$\Delta\le\sqrt{a(R)}\,e$.

For a finite training sample set
$F_m(p)=(2m)^{-1}\sum_i d(p,q_i)^2$. On this surface $\Hess F_m\succeq I$ and
$\nabla F_m=-m^{-1}\sum_i\Log_p(q_i)$ by the proof of
Theorem~\ref{thm:center}. If a certified gradient enclosure proves
$\|\nabla F_m(p)\|\le\epsilon$, strong monotonicity along the segment from $p$ to
the empirical minimizer gives $d(p,\hat\mu_m)\le\epsilon$. A centre certificate alone is not a
score certificate: errors in transport, log components, empirical covariance,
eigenvalue classification and rank decisions must also be propagated.

  \begin{figure*}[t]%
  \centering
  \includegraphics[width=\textwidth]{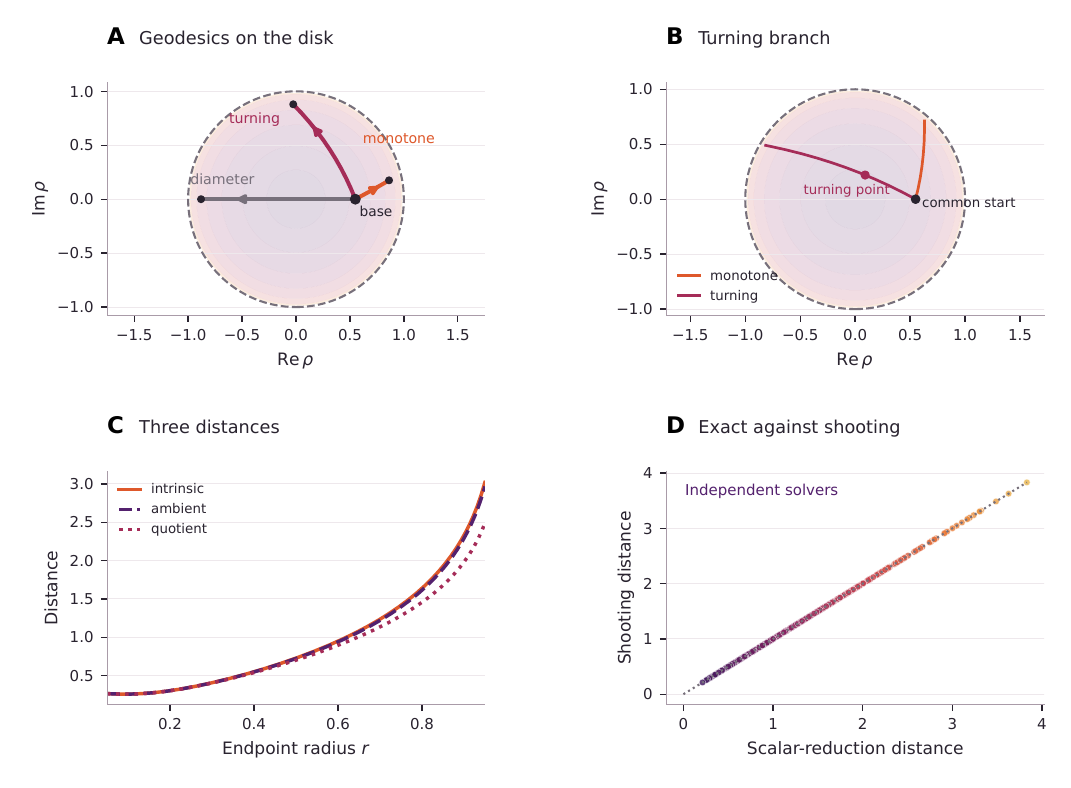}%
  \caption{\textbf{Geodesics admit a scalar reduction.} (A) Diameter, monotone-radius and turning paths from a common base point. (B) The minimum radius of a turning path, compared with a monotone path. Heat contours in A and B indicate intrinsic distance from the origin. (C) Intrinsic, ambient and quotient distances on identical endpoints. (D) Distance from the scalar reduction agrees with independent shooting of the Cartesian geodesic equation; color indicates distance.}\label{fig:figure_5.pdf}%
  \end{figure*}

\section{Estimation, calibration and independent-replicate inference}
\label{sec:calibration}

Every statement in this section refers to the fixed-complement reference law
$P$ of Section~\ref{sec:reference}, and uses the geometry proved in
Theorem~\ref{thm:hadamard}. Throughout, $X_1,\dots,X_m$ are independent with law
$P$, and $\hat\mu_m$ denotes their exact empirical intrinsic minimizer.

\subsection{Consistency}

\begin{theorem}[Consistency of the empirical centre]
\label{thm:consistency}
Under Eq.~\eqref{eq:secondmoment}, the empirical minimizer exists and is unique
almost surely, and $\hat\mu_m\to\mu$ almost surely.
\end{theorem}

\begin{proof}
Existence and uniqueness are Theorem~\ref{thm:center} applied to the empirical
measure. Write $F_{m,2}(y)=m^{-1}\sum_i d(y,X_i)^2$, so that the population
half-loss is $F$ and $F_{m,2}\to 2F$ pointwise by the strong law.

\emph{Localization.} Minimality of $\hat\mu_m$ together with the $L^2$ triangle
inequality gives
$d(o,\hat\mu_m)\le 2\bigl(F_{m,2}(o)\bigr)^{1/2}$, and the right side is bounded
eventually almost surely by the strong law. Choose the deterministic radius
$R_0>\max\{d(o,\mu),2(\mathbb E d(o,X)^2)^{1/2}\}$. The empirical
minimizers eventually lie in this ball almost surely.

\emph{Uniform law on that ball.} For $a,b$ in the ball,
\[
 \left|d(a,X)^2-d(b,X)^2\right|\le d(a,b)\left(2d(o,X)+2R_0\right)
\]
by the triangle inequality, and the right side is $P$-integrable. A finite net,
the strong law at the net points, and then a decreasing net mesh give
$\sup_{d(o,y)\le R_0}|F_{m,2}(y)-2F(y)|\to0$ almost surely.

\emph{Identification of the limit.} Any subsequential limit of $\hat\mu_m$ is a
minimizer of $F$ by the uniform law, and Theorem~\ref{thm:center} makes that
minimizer unique, so every convergent subsequence has limit $\mu$. Compactness
of the localization ball then gives convergence of the full sequence.
\end{proof}

The exact empirical centre satisfies the sample mean-logarithm equation by the
finite-sum first variation. A stopped numerical centre does not: it requires its
own error certificate, and a small reported gradient is not equality.

Covariance consistency follows by an explicit plug-in argument, and the estimated
centre is handled by continuity rather than left implicit. Fix a smooth
orthonormal frame near $\mu$ and let $v(y,x)$ denote log coordinates in that
frame. Assume continuity of $y\mapsto v(y,x)$ for almost every $x$ and an
envelope $B$ with $\sup_{y\in K}\|v(y,x)\|^2\le B(x)$ and $\mathbb E B(X)<\infty$
on a compact neighbourhood $K$ of $\mu$. These conditions hold here:
smoothness of the global logarithm gives continuity, and for
$K\subset\{y:d(y,\mu)\le R\}$ the envelope
$B(x)=2d(\mu,x)^2+2R^2$ is integrable by Eq.~\eqref{eq:secondmoment}.
Define the population second moment in
the same frame,
\begin{equation}
 \Sigma(y)=\mathbb E_P\,v(y,X)v(y,X)^\top ,
 \label{eq:popcov}
\end{equation}
with $\Sigma=\Sigma(\mu)$. For $y_k\to y$ the integrands converge pointwise by
continuity and are dominated by $B$, so dominated convergence gives
$\Sigma(y_k)\to\Sigma(y)$: the map $\Sigma(\cdot)$ is continuous on $K$, uniformly
in no further sense being needed. For $m\ge2$, define the empirical covariance
\begin{equation}
 \hat\Sigma_m=\frac{1}{m-1}\sum_{i=1}^m(\hat v_i-\bar v_m)(\hat v_i-\bar v_m)^\top,
 \qquad \hat v_i=v(\hat\mu_m,X_i),
 \label{eq:empcov}
\end{equation}
put $f_y(x)=v(y,x)v(y,x)^\top$ and
$M_\epsilon(x)=\sup_{d(y,\mu)\le\epsilon}\|f_y(x)-f_\mu(x)\|$, restricting the supremum to $K$.
Continuity gives $M_\epsilon(x)\to0$ pointwise, while $M_\epsilon\le2B$ gives
$\mathbb E M_\epsilon(X)\to0$ by dominated convergence. For each fixed $\epsilon$,
once $\hat\mu_m$ enters this ball,
\[
 \left\|\frac1m\sum_i f_{\hat\mu_m}(X_i)-\frac1m\sum_i f_\mu(X_i)\right\|
 \le\frac1m\sum_i M_\epsilon(X_i).
\]
The strong law on a countable sequence $\epsilon\downarrow0$ makes the left side
converge almost surely to zero. The supremum is measurable because it equals
the supremum on a countable dense subset of the compact parameter ball.
The fixed-centre second moment converges to
$\Sigma$ by the strong law. At the exact empirical centre,
$\bar v_m=0$ by the first-order equation, so the factor $m/(m-1)\to1$ yields
$\hat\Sigma_m\to\Sigma$ almost surely. Fitting the centre on the same data does
not imply that this estimator is unbiased.

\subsection{A central limit theorem on the induced surface}

The following is the chart estimating-equation argument of Bhattacharya and
Patrangenaru~\cite{S01,S02}, specialized to this surface with its derivative
conditions verified rather than assumed.

\begin{lemma}[Hessian comparison on the induced disk]
\label{lem:hessian}
Let $t=d(p,x)$ and $f_x(p)=\tfrac12d(p,x)^2$. In an orthonormal tangent frame
with the first axis along the geodesic from $p$ to $x$,
\begin{equation}
 I\preceq \Hess f_x(p)\preceq\left(1+\frac t2\right)I .
 \label{eq:hessian-bound}
\end{equation}
\end{lemma}

\begin{proof}
The radial eigenvalue is exactly one, because along the radial geodesic
$f_x=r^2/2$. For the transverse eigenvalue, let $J$ be the Jacobi field along the
unit-speed segment from $x$ to $p$ with $J(0)=0$ and $J(t)=v$, $|v|=1$. On a
surface, $\Hess f_x(v,v)=t\int_0^t(|J'|^2-KJ^2)\dd s$, the index form of $J$, and
$J$ minimizes it because there are no conjugate points. Since
$-1/4\le K\le0$, comparison with the spaces of constant curvature $-1/4$ and $0$
gives
\[
 1\ \le\ \Hess f_x(v,v)\ \le\ \frac t2\coth\frac t2 .
\]
The inequality $u\coth u\le1+u$ for $u>0$ (from
$\coth u=1+2/(\mathrm e^{2u}-1)\le1+1/u$) gives
$(t/2)\coth(t/2)\le1+t/2$. Radial and transverse cross terms vanish by the first
variation, so the bound holds in every direction. At $t=0$ the Hessian is the
identity and the bound is attained.
\end{proof}

\begin{theorem}[Sandwich central limit theorem]
\label{thm:clt}
Take normal coordinates $\theta$ at $\mu$, so $\theta_0=0$, and let
$\ell(\theta,x)=d(\exp_\mu\theta,x)^2/2$ with
$G=\partial_\theta\ell$ and $H=\partial_\theta G$. Assume that exact consistent
empirical minimizers lie eventually in the chart interior, that $\ell$ is twice
continuously differentiable there for almost every $x$, that
$\mathbb E\|G(0,X)\|^2<\infty$ and $\mathbb E\|H(0,X)\|<\infty$, that
\begin{equation}
 \mathbb E\sup_{\|\theta\|\le\epsilon}\|H(\theta,X)-H(0,X)\|\longrightarrow0
 \quad(\epsilon\downarrow0),
 \label{eq:hesscont}
\end{equation}
and that $A=\mathbb E H(0,X)$ is nonsingular. Then
\begin{equation}
 \begin{gathered}
 \sqrt m\,\hat\theta_m\ \Longrightarrow\ N\!\left(0,\ A^{-1}\Sigma A^{-\top}\right),\\
 \Sigma=\mathbb E\,G(0,X)G(0,X)^\top .
 \end{gathered}
 \label{eq:sandwich}
\end{equation}
\end{theorem}

\begin{proof}
Taylor expansion in integral form about $\theta=0$ gives, at the empirical
minimizer,
\begin{gather*}
 0=\frac1{\sqrt m}\sum_{i=1}^mG(0,X_i)+A_m\sqrt m\,\hat\theta_m,\\
 A_m=\frac1m\sum_{i=1}^m\int_0^1H(t\hat\theta_m,X_i)\,\dd t .
\end{gather*}
To handle the dependence of $\hat\theta_m$ on the same sample, put
$B_\epsilon(x)=\sup_{\|\theta\|\le\epsilon}\|H(\theta,x)-H(0,x)\|$.
On $\{\|\hat\theta_m\|\le\epsilon\}$,
$\|A_m-m^{-1}\sum_iH(0,X_i)\|\le m^{-1}\sum_iB_\epsilon(X_i)$.
For any $\eta>0$, Markov's inequality bounds the limsup probability that
this norm exceeds $\eta$ by $\mathbb E B_\epsilon(X)/\eta$, since
consistency makes the complementary event vanish. Letting
$\epsilon\downarrow0$ and using Eq.~\eqref{eq:hesscont}, followed by the
strong law for $H(0,X)$, proves $A_m\to A$ in probability. At $\theta=0$ the gradient is $G(0,X)=-\Log_\mu X$ by
Theorem~\ref{thm:center}, so $\mathbb E G(0,X)=0$ and
$\Cov G(0,X)=\Sigma$. The multivariate central limit theorem for independent
identically distributed summands gives
$m^{-1/2}\sum_iG(0,X_i)\Rightarrow N(0,\Sigma)$, and Slutsky's theorem with
$A_m\to A$ gives Eq.~\eqref{eq:sandwich}.
\end{proof}

Using $d^2$ instead of $d^2/2$ multiplies $G$ and $H$ by two and $\Sigma$ by four,
leaving the sandwich unchanged. The limit is not $\Sigma/m$ in general: curvature
enters through $A$. Lemma~\ref{lem:hessian} supplies Eq.~\eqref{eq:hesscont} for
this surface, because on a compact normal-coordinate ball the chart Hessian is
bounded by $C(1+d(\mu,x))$ with $C$ depending only on the ball and
$\mathbb E d(\mu,X)\le(\mathbb E d^2)^{1/2}<\infty$; see
Appendix~\ref{app:moments}. At the centre $A\succeq I$, so $A$ is finite and
nonsingular. On the real arc, $A=1$ exactly.

No conclusion about individual residuals follows. Equation~\eqref{eq:sandwich}
describes the sampling distribution of an estimated centre; it does not make an
individual observation Gaussian, and it does not make the joint score of
Remark~\ref{rem:notgaussian} chi-square.

\begin{corollary}[Accuracy required of a numerical centre]
\label{cor:numerical-centre}
Let $F_m(y)=(2m)^{-1}\sum_i d(y,X_i)^2$.
If a computed centre $\tilde\mu_m$ satisfies an exact residual bound
$\|\nabla F_m(\tilde\mu_m)\|\le\epsilon_m$, then
$d(\tilde\mu_m,\hat\mu_m)\le\epsilon_m$. Thus $\epsilon_m\to0$ almost surely
preserves strong consistency, and $\sqrt m\,\epsilon_m\to0$ in probability preserves the limit in
Eq.~\eqref{eq:sandwich}.
\end{corollary}
\begin{proof}
Strong convexity gives
$\|\nabla F_m(y)\|\ge d(y,\hat\mu_m)$ by integrating its Hessian along
the geodesic from the minimizer. The logarithm at $\mu$ is one-Lipschitz
on a Hadamard manifold, so the normal-coordinate difference between the
computed and exact centres is at most $\epsilon_m$. The conclusions follow
from consistency and Slutsky's theorem.
\end{proof}
A fixed floating-point stopping threshold is a finite-sample diagnostic.
Applying this corollary requires a bound on the exact residual, including
errors in the computed logarithms.

\subsection{Frozen scores and the finite-sample rank}

In practice a score is a frozen numerical procedure fitted on training data, not
an ideal geometric quantity. The following theorem is stated for such a procedure
and needs only measurability and exchangeability.

Let $\mathcal T$ denote the training data together with every deterministic
choice made from them: the centre, the frame, the covariance, the spectral
thresholds, the retained projector $\hat P$ and the pseudoinverse
$\hat\Sigma^+$. These define one frozen measurable function $s_{\mathcal T}$ on
the slice.

For example, fix an eigenvalue threshold from training and retain the
corresponding eigenspace. For a new tangent vector $v$, report
$v^\top\hat\Sigma^+v$ on that space and the residual
$\|(I-\hat P)v\|$ separately. A scalar calibration rule may assign
$+\infty$ when this residual exceeds a tolerance fixed at training time.
Use the identical rule for each calibration and test point. At rank zero
only the support-departure rule has a meaning. A failed logarithm or
centre fit supplies no score.

\begin{theorem}[Finite calibration rank]
\label{thm:rank}
Conditional on $(\mathcal T,c)$, where $c$ denotes the declared complement,
assume that $n_{\mathrm{cal}}$ calibration
observations and one null test observation are exchangeable with the same law, and
that the frozen measurable score is defined for all of them. Write $s_i$ for the
calibration scores and $s_*$ for the score of the null test observation. Then,
with $N_c=n_{\mathrm{cal}}+1$,
\begin{equation}
 p=\frac{1+\sum_{i=1}^{n_{\mathrm{cal}}}\mathbf 1\{s_i\ge s_*\}}{N_c}
 \label{eq:rankp}
\end{equation}
is super-uniform:
\begin{equation}
 \Pr(p\le a\mid\mathcal T,c)\le\frac{\lfloor aN_c\rfloor}{N_c}\le a,
 \qquad 0\le a\le1 .
 \label{eq:superuniform}
\end{equation}
\end{theorem}

\begin{proof}
For the multiset of $N_c$ scores put $R_j=\#\{i:s_i\ge s_j\}$ and arrange the
tie blocks in descending order, including a top block of $+\infty$ values when
present. In a block of size $b$ with $h$ strictly higher observations, every
member has $R_j=h+b$. Hence among all indices at most $k$ satisfy $R_j\le k$:
such blocks can only end within the first $k$ positions of the ordering.
Exchangeability makes the test index uniform over the $N_c$ positions
conditional on the unordered score multiset, so
$\Pr(R_*\le k\mid\mathcal T,c)\le k/N_c$. Since $p\le a$ implies
$R_*\le\lfloor aN_c\rfloor$, Eq.~\eqref{eq:superuniform} follows. Ties are
included by the relation $\ge$, and if all scores equal $+\infty$ then $p=1$.
\end{proof}

This is the split-rank argument underlying the split-conformal
principle~\cite{S10}. The guarantee averages over the calibration sample and the
new null, conditional on the fit and the complement; it is not conditional on
realized calibration scores. Shared calibration makes many test ranks dependent
without spoiling each marginal guarantee. Bonferroni over valid marginal ranks
controls familywise error by the union bound; no automatic Benjamini--Hochberg
claim follows.

The theorem applies to any defined common computational function, even one that
differs from the ideal geometry. It does not certify numerical accuracy of the
centre, logarithm, covariance or quadratic score. Solver tolerances and
refinement agreement are separate diagnostics. A computational support infinity
is an output definition, whereas an arithmetic failure is unavailable and must
not be relabelled as infinity; if the procedure is partial on the reference law,
the total-function hypothesis has not been established, and selecting only
successful computations cannot restore the unconditional guarantee.

\subsection{Independent replicates and multiplicity}
For independent planned clusters, bounded-variable concentration gives
conservative uncertainty intervals even when observations within a cluster
are dependent. Holm's step-down rule controls familywise error for a family
of valid marginal equality tests without requiring independence between
their ranks~\cite{S21,S22}. Supplementary Methods S8 gives the bounds and
their derivations, including paired comparisons and unavailable outcomes.
These guarantees use complete planned clusters as the independent units.

\subsection{Varying complements}

The preceding theory conditions on a fixed complement. When the complement is
itself estimated, the normalized connection inherits derivatives from every
surrounding entry; Appendix~\ref{app:varying} gives the exact differential. Two
consequences matter here. First, a fixed population matrix used for oracle
normalization does not make the sample complement fixed, and oracle coordinates
can leave the admissible disk, so no conditional slice calibration guarantee is
available in that regime. Second, if a matrix estimator satisfies
$\sqrt m(\mathbf a_m-\mathbf a_0)\Rightarrow N(0,\Omega)$ in a fixed real
vectorization, then for any smooth real functional $\varphi$ of the normalized
connection the delta method gives
$\sqrt m[\varphi(\mathbf a_m)-\varphi(\mathbf a_0)]
\Rightarrow N(0,\nabla\varphi^\top\Omega\nabla\varphi)$, provided the limiting
variance is positive and consistently estimated. A standardized normal limit
requires those two additional conditions and is not a consequence of smoothness
alone.

  \begin{figure*}[t]%
  \centering
  \includegraphics[width=\textwidth]{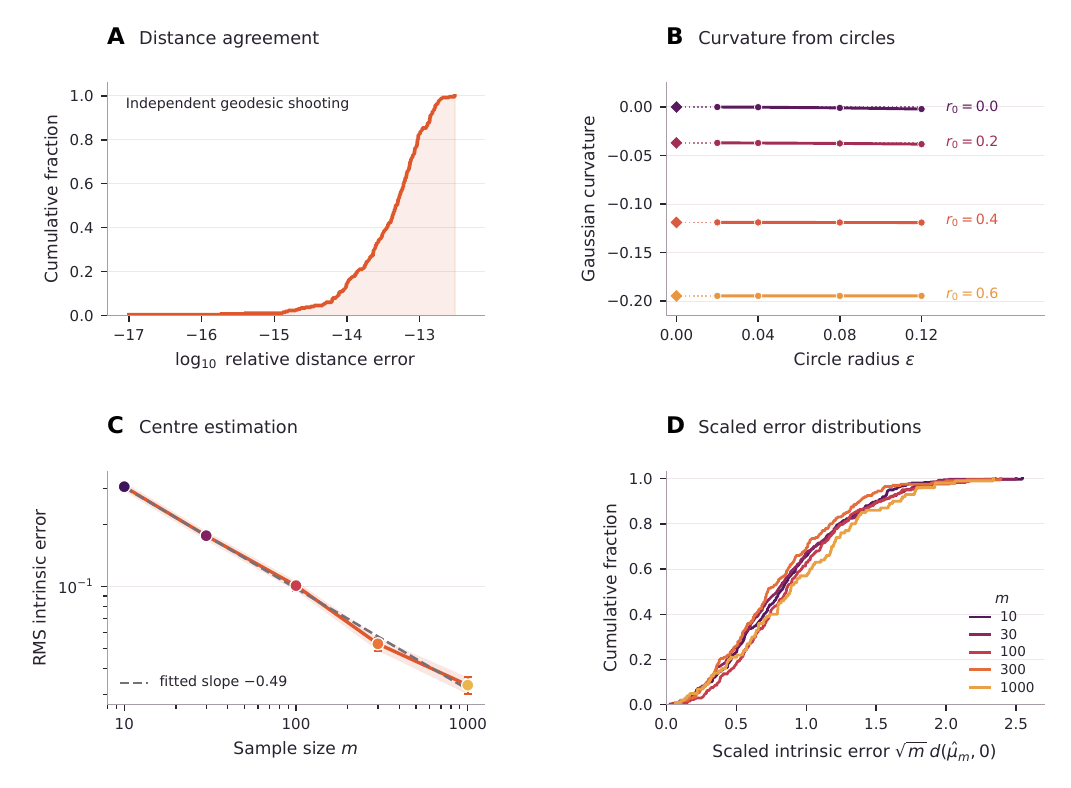}%
  \caption{\textbf{Numerical accuracy and intrinsic centre estimation.} (A) Cumulative relative distance error against independent shooting; values below $10^{-17}$ are placed at that plotting floor. (B) Curvature recovered from geodesic circles approaches the exact values, marked at zero circle radius. (C) RMS centre error with pointwise 95\% bootstrap intervals and a fitted logarithmic slope. (D) Empirical distributions of $\sqrt{m}\,d(\hat\mu_m,0)$ across sample sizes. Sample counts, solver settings and numerical tables are provided in the Supplementary Methods.}\label{fig:figure_6.pdf}%
  \end{figure*}

\section{Numerical illustrations}
\label{sec:numerical}

The numerical examples examine the geodesic formulas and illustrate intrinsic
statistics under declared reference laws. The Supplementary Methods give the
sampling designs, solver tolerances, replication counts and numerical tables.

\subsection{Geodesics and curvature}

Independent shooting of the Cartesian geodesic equation agrees with the scalar
distance reduction over the tested interior range
(Figures~\ref{fig:figure_5.pdf}D and~\ref{fig:figure_6.pdf}A). The largest relative
discrepancy is $3.05\times10^{-13}$. Shooting starts from a Euclidean initial
guess and evaluates length through the metric norm of its fitted initial
velocity, independently of the scalar formula.

Curvature estimates from geodesic-circle circumferences approach the values of
Theorem~\ref{thm:hadamard} as the circle radius decreases
(Figure~\ref{fig:figure_6.pdf}B). The finite-radius correction is most visible
at the origin, where the limiting curvature is zero. Together, these
comparisons assess the implementation on an interior domain; the behaviour
arbitrarily close to the singular boundary requires the numerical
representations discussed in Appendix~\ref{app:stablepsi}.

\subsection{Intrinsic statistics}

For the centre experiment, a rotationally symmetric Gaussian law is mapped to
the disk by $Y\mapsto\tanh(\|Y\|)Y/\|Y\|$. Its population centre is the origin,
and Appendix~\ref{app:moments} verifies the finite second intrinsic moment.
The empirical centre error is compatible with the $m^{-1/2}$ scaling of
Theorem~\ref{thm:clt}: the fitted logarithmic slope is $-0.490$, with a
95\% bootstrap interval $[-0.511,-0.471]$
(Figure~\ref{fig:figure_6.pdf}C). Distributions of the scaled errors show the
associated finite-sample variation (Figure~\ref{fig:figure_6.pdf}D).

Figure~\ref{fig:figure_3.pdf} separates the roles of population geometry and
score calibration. A Gaussian tangent law illustrates covariance whitening and
rotations that preserve the covariance spectrum. For calibration, a training
sample from a uniform disk supplies a fitted centre and covariance, which are
then held fixed. Independent calibration and test samples evaluate the
resulting quadratic score. Rounding retains ties and illustrates the
conservative rank rule under the exchangeable design of
Theorem~\ref{thm:rank}. The plotted uncertainty intervals quantify Monte Carlo
variation conditional on the training fit.

For a covariance-level comparison, four-point tangent laws give known centres
and allow direct evaluation of the population sandwich matrix. Across the
three laws and two sample sizes in Supplementary Methods S9, relative
covariance discrepancies ranged from 0.065 to 0.164. Coverage of nominal
95\% ellipses constructed with the population sandwich ranged from 0.938 to
0.963. These experiments assess the covariance prediction with known
population parameters. Practical ellipses estimated from a single training
sample introduce additional uncertainty.

\section{Discussion}
\label{sec:discussion}

Fixing the surrounding entries of a Hermitian positive-definite matrix gives an
isolated connection a complete, two-dimensional geometry. The fixed Schur
congruence identifies its admissible disk and removes the ambient dimension from
the induced tangent metric. The variable nonpositive curvature guarantees unique
geodesics, global logarithms and a unique reference centre under a finite second
intrinsic moment. Conservation of angular momentum then reduces geodesic
computation to one scalar equation with elementary angular and length integrals.

The constraint determines the meaning of the distance. Ambient affine-invariant
paths can change diagonal entries, whereas quotient paths identify matrices under
diagonal rescaling. On the fixed-complement slice, distinct endpoints have a
strictly larger intrinsic distance than their ambient distance. Correlation
quotients and transformed anomaly scores provide the established background
\cite{G01,G02,G03,G04}. The present construction specifies the path constraint
explicitly and supplies its exact metric and geodesic reduction. The real
restriction also retains an orientation through its signed arc coordinate.
Log-Euclidean and log-Cholesky constructions provide alternative coordinate
geometries~\cite{G16,G20,G21} whose paths answer different questions.

The statistical interpretation requires a reference distribution on the declared
slice. A common normalized geometry across slices does not identify a common
probability law. Fr\'echet means, tangent covariance and the sandwich limit give
population summaries and uncertainty for an estimated centre. Covariance
whitening fixes the first two tangent moments but leaves the individual score
distribution dependent on the reference law. Exchangeable rank calibration
provides a finite-sample guarantee for a fitted scoring rule when the training,
calibration and test design meets the stated assumptions. The numerical examples
illustrate these distinctions with a declared Gaussian tangent law and an
independent calibration experiment.

The framework applies directly to controlled connection perturbations with fixed
surrounding entries. Covariance estimates from different subjects or samples
generally have varying complements. Their normalized coordinates inherit
variation from all matrix entries, and their sampling laws require a separate
argument. Appendix~\ref{app:varying} gives the relevant differential and a
delta-method limit under a matrix-estimator central limit theorem. The
slice-preserving invariances likewise depend on the declared frame and on
transporting the reference law consistently. Generic variable mixing changes the
isolated-entry problem.

Numerical evaluation becomes sensitive near the excluded boundary, where
intrinsic distance diverges. Stable arc evaluation and explicit solver status
checks are therefore part of the implementation. The reported shooting and
curvature checks compare numerical evaluations with the derived formulas over a
specified interior range. Extending the representation through log-gap
coordinates and validating error enclosures would support computations closer
to singular matrices. Applied comparisons of fixed-complement scores with other
matrix summaries require a sampling design matched to the intended population.

\section*{Acknowledgements}
This work was supported in part by the STI 2030--Major Projects (2022ZD0208500);
the National Natural Science Foundation of China (W2411084); the Key Research and
Development Projects of the Science and Technology Department of Chengdu
(2024-YF08-00072-GX); and the Chengdu Science and Technology Bureau Program
(2022-GH02-00042-HZ). Additional support was provided by the University of
Electronic Science and Technology of China through the CNS Program
(Y0301902610100201).

\section*{Competing interests}
The authors declare no competing interests.

\section*{Code availability}
The implementation and scripts for reproducing the numerical examples are
maintained at \url{https://github.com/ronald1129/sid-hpd-intrinsic}.
Repository access is available to reviewers on request.

\begingroup
\small
\bibliographystyle{plainnat}
\bibliography{references}
\endgroup

\clearpage
\appendix
\setcounter{figure}{0}
\renewcommand{\thefigure}{A\arabic{figure}}
  \twocolumn[{\begin{minipage}{\textwidth}%
  \centering\includegraphics[width=\textwidth]{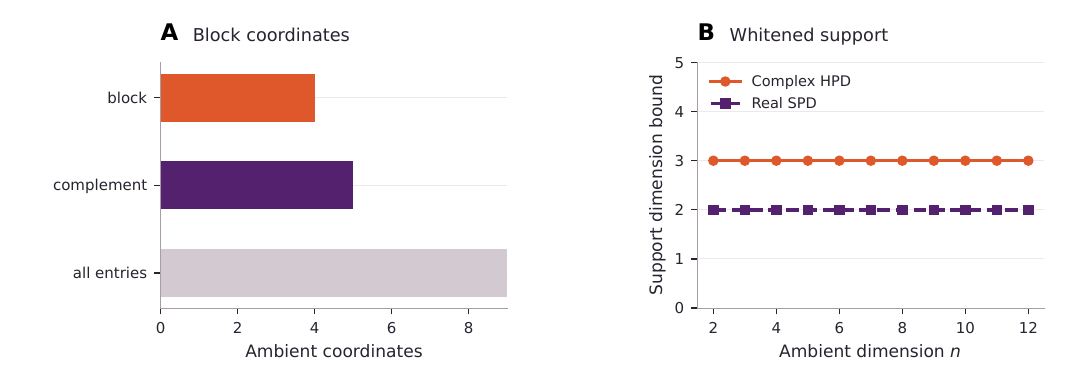}%
  \captionof{figure}{\textbf{Block storage and ambient logarithm support are distinct.} (A) Real coordinate counts for a Hermitian matrix of order three and its selected block. (B) Conjugate-swap symmetry confines the whitened ambient logarithms to at most three coordinates in the complex case and two in the real case, independently of matrix dimension. These support bounds refine the four- and three-coordinate block storage representations of Proposition~\ref{prop:ambientsupport}.}\label{fig:figure_s1.pdf}%
  \vspace{10pt}\end{minipage}}]

\section{Exact ambient support reduction}
\label{app:ambient}

The fixed Schur congruence of Theorem~\ref{thm:reduction} also gives a
computational reduction for the unrestricted ambient comparator.

\begin{proposition}[The ambient comparator lives in a two-by-two block]
\label{prop:ambientsupport}
Let $P$ be a fixed-complement law with finite ambient second moment. In the
coordinates supplied by the congruence, the unrestricted ambient Fr\'echet mean of
$P$ is $\diag(M_2,I_{n-2})$ with $M_2$ the HPD$(2)$ ambient mean of the two-by-two
blocks. Every whitened ambient logarithm has only a Hermitian two-by-two block, so
full ambient tangent covariance standardization is isometrically representable in
four block coordinates (three in the real SPD case), independently of $n$.
The fixed slice has an additional symmetry: the whitened logarithms actually
lie in a linear subspace of dimension at most three in the complex case and
at most two in the real case. Both reductions retain the complete ambient
tangent vector and its covariance quadratic score.
\end{proposition}

\begin{proof}
Let $J=\diag(I_2,-I_{n-2})$. The map $A\mapsto JAJ$ is an isometry of the ambient
manifold with Eq.~\eqref{eq:airm} and fixes every observation of the slice, since
those observations equal $\diag(R(\rho),I_{n-2})$ in these coordinates. A unique
ambient Fr\'echet mean commutes with any isometry fixing the data, so the mean is
block diagonal.

For a block-diagonal candidate $\diag(B,D)$ its squared distance to an observation
splits as $d_A(B,R)^2+d_A(D,I)^2$, because the matrix logarithm of a block
diagonal matrix is block diagonal. Minimizing the second term forces $D=I$, and the
first term is minimized by $B=M_2$, the HPD$(2)$ ambient mean. The same block
formula gives whitened logarithms
$\diag\bigl(\log(M_2^{-1/2}RM_2^{-1/2}),0\bigr)$.

For the sharper support bound, write
$S=\begin{psmallmatrix}0&1\\1&0\end{psmallmatrix}$ and consider the isometry
$\iota(B)=S\bar B S$ of HPD$(2)$. Every $R(\rho)$ is fixed by $\iota$.
Uniqueness of the ambient mean gives $\iota(M_2)=M_2$. This involution
commutes with products and real spectral matrix functions; hence each
$H_\rho=\log(M_2^{-1/2}R(\rho)M_2^{-1/2})$ is fixed by $\iota$ as well.
A Hermitian fixed point has the form
$H_\rho=\begin{psmallmatrix}t&u+\mathrm iv\\u-\mathrm iv&t\end{psmallmatrix}$,
with $t,u,v\in\R$. Its Frobenius norm is the Euclidean norm of
$\sqrt2(t,u,v)$, proving the three-coordinate representation. In the real
case $v=0$, giving two coordinates. The actual covariance rank may be
smaller, depending on the reference law.

Hermitian vectorization preserves Frobenius inner products and therefore
covariance eigenvalues, pseudoinverse quadratic scores and the support-departure
norm. Under the original congruence, if $M'=TMT^*$ then
$O=M'^{-1/2}TM^{1/2}$ is unitary and whitened logarithms transform by
$H\mapsto OHO^*$, which is an orthogonal change of whitened coordinates. The
reduction therefore preserves the original full ambient comparator.
\end{proof}

This reduction does not replace whole-matrix scores by an edge projection: it is
valid only because every observation in a replication shares one fixed complement.
It does not apply to the varying-complement regime of
Appendix~\ref{app:varying}, and it cannot manufacture full-rank population
covariance from a singular empirical covariance. It also does not imply that the
HPD$(2)$ ambient mean has unit diagonals, as the two-point counterexample of
Section~\ref{sec:comparison} shows.
Figure~\ref{fig:figure_s1.pdf} records the retained-coordinate counts before and
after the congruence.

  \begin{figure*}[t]%
  \centering
  \includegraphics[width=\textwidth]{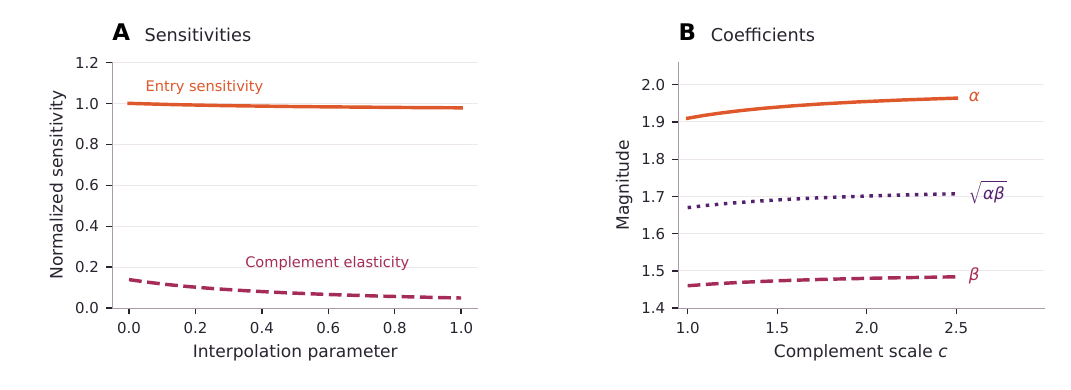}%
  \caption{\textbf{The complement affects both sensitivity and normalization.} (A) Normalized sensitivity to the selected entry and elasticity with respect to the complement scale. (B) The coefficients $\alpha$, $\beta$ and $\sqrt{\alpha\beta}$ along the same declared interpolation. Both panels evaluate the differential formulas in Appendix~\ref{app:varying}; only the complement scale varies.}\label{fig:figure_s2.pdf}%
  \end{figure*}

\section{Differentiating the normalization when the complement varies}
\label{app:varying}

Geometric conditioning and covariance-estimator sampling are different regimes,
and Section~\ref{sec:calibration} uses the transition between them. For an interior
HPD matrix let $H=EC^{-1}E^*$ and let $u,v$ denote the first and second rows of
$E$. A Hermitian variation gives, by
$\dd C^{-1}=-C^{-1}(\dd C)C^{-1}$ and the product rule,
\begin{align}
 \dd H={}&(\dd E)C^{-1}E^*+EC^{-1}\dd E^*
   -EC^{-1}(\dd C)C^{-1}E^*, \label{eq:dH}\\
 \dd\alpha={}&\dd a-2\operatorname{Re}\!\left[(\dd u)C^{-1}u^*\right]
   +uC^{-1}(\dd C)C^{-1}u^*, \label{eq:dalpha}\\
 \dd\beta={}&\dd b-2\operatorname{Re}\!\left[(\dd v)C^{-1}v^*\right]
   +vC^{-1}(\dd C)C^{-1}v^*, \label{eq:dbeta}\\
 \dd\gamma={}&(\dd u)C^{-1}v^*+uC^{-1}\dd v^*
   -uC^{-1}(\dd C)C^{-1}v^*, \label{eq:dgamma}\\
 \dd\rho={}&\frac{\dd z-\dd\gamma}{\sqrt{\alpha\beta}}
   -\frac{\rho}{2}\left(\frac{\dd\alpha}{\alpha}
   +\frac{\dd\beta}{\beta}\right). \label{eq:drho}
\end{align}
For $n=2$ the terms involving $H$ are deleted. These expressions are real
derivatives of the complex normalized coordinate, obtained by differentiating
Eq.~\eqref{eq:schur} and Definition~\ref{def:slice} entrywise.

For real SPD sampling, $\varphi(A)=\psi(\rho(A))$ is smooth on the interior. If a
matrix estimator satisfies $\sqrt m(\mathbf a_m-\mathbf a_0)\Rightarrow
N(0,\Omega)$ in a fixed real vectorization, then the delta method gives
\begin{equation}
 \sqrt m\left[\varphi(\mathbf a_m)-\varphi(\mathbf a_0)\right]
 \Rightarrow N\!\left(0,\ \nabla\varphi^\top\Omega\,\nabla\varphi\right),
 \label{eq:deltareal}
\end{equation}
provided the limiting variance is positive and consistently estimated. For complex
sampling one uses a declared smooth two-component function, for example an
intrinsic logarithm at a fixed interior reference followed by its orthonormal
coordinates; its Jacobian supplies a multivariate delta-method limit. No canonical
signed complex scalar is implied. If the reference centre is itself estimated, its
derivative and dependence must be included in the joint limit rather than held
fixed silently.

For centred real Gaussian data the sample covariance fluctuation has entries
$\Cov(\sqrt m\,S_{ab},\sqrt m\,S_{cd})\to M_{ac}M_{bd}+M_{ad}M_{bc}$. For proper
complex Gaussian data, writing $S=m^{-1}\sum\mathbf x\mathbf x^*$ before the
asymptotically negligible centring adjustment,
\begin{align}
 \mathbb E\left[\delta S_{ab}\overline{\delta S_{cd}}\right]&=m^{-1}M_{ac}M_{db},
 \label{eq:cwishart1}\\
 \mathbb E\left[\delta S_{ab}\,\delta S_{cd}\right]&=m^{-1}M_{ad}M_{cb},
 \label{eq:cwishart2}
\end{align}
where $\delta S=S-M$. These follow by pairing Gaussian fourth moments and
subtracting the product of means; Hermitian redundancy must be respected when
assembling a real covariance matrix. They are sampling covariances, not intrinsic
individual reference laws. A fixed population matrix used for oracle normalization
does not make the sample complement fixed, and oracle coordinates can leave the
admissible disk; such failures are retained and receive no conditional slice
guarantee. Figure~\ref{fig:figure_s2.pdf} shows how the normalized coordinate
inherits the slower of the two dependencies along an interpolation of the
complement.

  \begin{figure*}[t]%
  \centering
  \includegraphics[width=\textwidth]{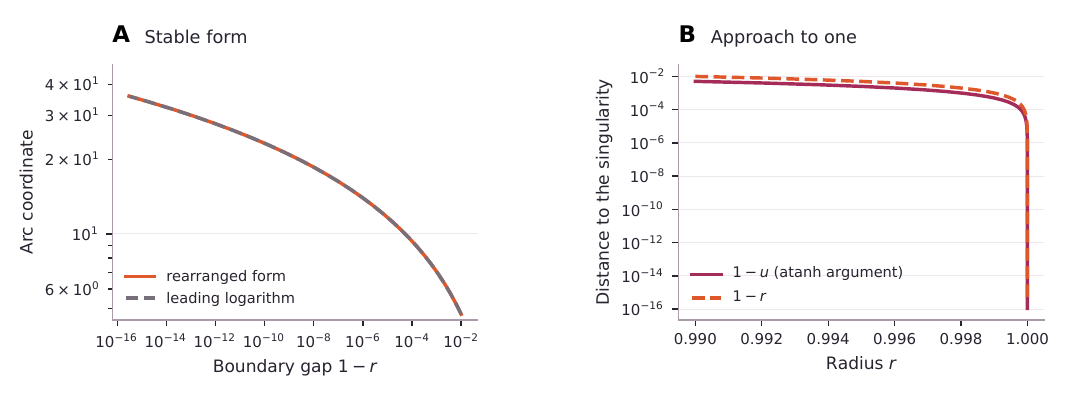}%
  \caption{\textbf{Stable arc evaluation near the excluded boundary.} (A) The rearranged arc coordinate and its leading logarithm as the boundary gap vanishes. (B) The gaps $1-u$ and $1-r$, where $u=\sqrt2r/\sqrt{1+r^2}$ is the direct inverse-hyperbolic-tangent argument. Keeping the boundary gap explicit avoids evaluating that function at a rounded argument of one.}\label{fig:figure_s3.pdf}%
  \end{figure*}

\section{Stable evaluation of the arc coordinate}
\label{app:stablepsi}

The closed form~\eqref{eq:psi} should not be evaluated by taking the inverse
hyperbolic tangent of a number that has already been rounded to one. For
$0\le r<1$ put
\begin{equation}
 u=\frac{\sqrt2\,r}{\sqrt{1+r^2}},\qquad \Delta=(1-r)(1+r),
 \label{eq:ustable}
\end{equation}
so that $1-u^2=\Delta/(1+r^2)$ by direct expansion. Since
$2\operatorname{atanh}u=\log(1+u)-\log(1-u)$ and
$\log(1-u^2)=\log(1+u)+\log(1-u)$, one has
$2\operatorname{atanh}u=2\log(1+u)-\log(1-u^2)$, and therefore
\begin{equation}
 \psi(r)=2\log(1+u)+\log(1+r^2)-\log\Delta-\sqrt2\operatorname{asinh}r .
 \label{eq:psistable}
\end{equation}
Every term in Eq.~\eqref{eq:psistable} is finite for $0\le r<1$; only
$-\log\Delta$ grows, and it grows exactly as the boundary is approached, which is
the divergence identified in Theorem~\ref{thm:hadamard}. Evaluating with
log-one-plus for small arguments and using the odd symmetry of $\psi$ for negative
real coordinates avoids the cancellation that affects the original form.

Near the origin a series or an enclosed quadrature is preferable when subtraction
loss matters. The boundary gap must be represented rather than silently clipped:
if an input has already been rounded to $r=1$, the admissible finite point is no
longer identified by that floating-point value. Retaining a log gap or higher
precision preserves the declared latent input and requires an explicit
conversion policy between numerical representations. The implementation rejects an inverse arc value when its residual cannot be resolved in the chosen floating-point representation. Figure~\ref{fig:figure_s3.pdf} displays the
rearranged form against the boundary gap together with its leading logarithm.

\section{Moment conditions and the verification of the central limit theorem}
\label{app:moments}

\subsection{Bounded chart derivatives}

Theorem~\ref{thm:clt} assumes that the chart Hessian is locally uniformly
continuous with an integrable envelope. Lemma~\ref{lem:hessian} supplies the
required bound.

On a compact normal-coordinate ball $B$ about $\mu$, the metric coefficients,
their first derivatives and the Christoffel symbols are bounded by constants that
depend only on $B$. The ordinary coordinate Hessian of $\ell$ is the covariant
Hessian plus a connection-coefficient term times the gradient, and the gradient
has norm $d(p,x)$. Lemma~\ref{lem:hessian} therefore gives
\begin{equation}
 \|\Hess_\theta\ell(\theta,x)\|\le C_B\left(1+d(\mu,x)\right)
 \label{eq:charthess}
\end{equation}
uniformly for $\theta$ in the ball, with $C_B$ depending only on $B$. The loss is
smooth in $p$ for every $x$ because the global logarithm of
Theorem~\ref{thm:hadamard} has no cut locus. Pointwise continuity of the
coordinate Hessian, together with the envelope~\eqref{eq:charthess} and
$\mathbb E d(\mu,X)\le(\mathbb E d^2)^{1/2}<\infty$, verifies the local uniform
continuity condition~\eqref{eq:hesscont} by dominated convergence.

At the centre the expected Hessian satisfies $A\succeq I$ by
Lemma~\ref{lem:hessian}, so $A$ is finite and nonsingular. Consistency and the
sandwich limit of Theorem~\ref{thm:clt} therefore hold on this disk for
independent identically distributed observations with finite second intrinsic
moment. On the real arc $A=1$ exactly.

\subsection{Declared generating laws}

\begin{lemma}[Intrinsic radius versus Euclidean radius]
\label{lem:latentmoment}
For the latent sampler $\rho=\tanh(\|Y\|)Y/\|Y\|$, with value zero at $Y=0$,
\begin{equation}
 \sqrt2\,\|Y\|\ \le\ d(0,\rho)=\psi\bigl(\tanh\|Y\|\bigr)\ \le\ 2\|Y\| .
 \label{eq:latentmoment}
\end{equation}
\end{lemma}

\begin{proof}
Write $v=\|Y\|$ and differentiate:
\begin{align*}
 \frac{\dd}{\dd v}\psi(\tanh v)
 &=\psi'(\tanh v)\operatorname{sech}^2v\\
 &=\frac{\sqrt2\sqrt{1+\tanh^2v}}{1-\tanh^2v}\left(1-\tanh^2v\right)\\
 &=\sqrt2\sqrt{1+\tanh^2v},
\end{align*}
using $\psi'(r)=\sqrt2\sqrt{1+r^2}/(1-r^2)$ from
Proposition~\ref{prop:psi}. Since $0\le\tanh^2v\le1$, the derivative lies in
$[\sqrt2,2]$; integrating from $0$ and using $\psi(0)=0$ gives
Eq.~\eqref{eq:latentmoment}.
\end{proof}

Consequently Eq.~\eqref{eq:secondmoment} holds for every latent law with finite
second Euclidean moment. Gaussian laws, finite Gaussian mixtures and centred
laws with exponential components all qualify. A multivariate $t_\nu$ law has
radial tail density proportional to $v^{-\nu-1}$, so its second moment is finite if and only if $\nu>2$. A $t_5$ law satisfies
Eq.~\eqref{eq:secondmoment}; a $t_1$ law does not. For the latter, the finite
second-moment centre, covariance and central-limit construction used here does
not apply; this does not exclude other notions of location.
A near-boundary family with $r\le1-10^{-5}$ has bounded intrinsic distance and
therefore also satisfies Eq.~\eqref{eq:secondmoment}.

These are analytic checks of the declared generators, not measured outcomes. No
corresponding moment assertion is made for the varying-complement oracle mapping
of Appendix~\ref{app:varying}, which can be undefined outside the admissible disk,
and none is made for a law whose second intrinsic moment has not been verified.

\end{document}